\documentclass[letterpaper,twocolumn,aps,pra,floatfix,superscriptaddress,footinbib,preprintnumbers]{revtex4-2}
\pdfoutput=1
\usepackage[utf8]{inputenc}
\usepackage{amsmath}
\usepackage{amssymb}
\usepackage{amsfonts}
\usepackage{amsthm}
\usepackage{physics}
\usepackage{orcidlink}
\usepackage{hyperref}
\usepackage{mathtools}
\usepackage{lipsum}
\usepackage{verbatim}
\usepackage{url}

\newcommand{\JQI}{Joint Quantum Institute, National Institute of Standards and Technology and
  University of Maryland, College Park, Maryland 20742, USA}

\begin{document}
\title{High-Fidelity Microwave-Polarization Control in a Rydberg-Ensemble Experiment}

\author{Deniz Kurdak\,\orcidlink{0000-0003-4076-3013}}
\email{dkurdak@umd.edu}
\affiliation{\JQI}
\author{Yaxin Li\,\orcidlink{0000-0001-8734-0136}}
\affiliation{\JQI}
\author{Patrick R. Banner\,\orcidlink{0009-0006-9957-4996}}
\affiliation{\JQI}
\author{J. V. Porto\,\orcidlink{0000-0002-6290-7535}}
\affiliation{\JQI}
\author{S. L. Rolston\,\orcidlink{0000-0003-1671-4190}}
\affiliation{\JQI}


\date{\today}

\begin{abstract}
Control of the polarization of microwave fields is a key experimental capability for a number of atomic physics platforms. 
However, producing high-fidelity microwaves requires a well-controlled microwave environment, where reflections that distort the polarization must be avoided or well characterized, a constraint that often conflicts with other experimental design considerations.
Here we demonstrate a microwave control system capable of producing high-fidelity microwave polarizations in a Rydberg-ensemble experiment. 
We use three in-vacuum DC electrodes, repurposed as microwave antennae, to produce imperfect and initially unknown polarizations. Each source is driven with independent phase and amplitude control to generate the desired microwave fields. We probe the fields produced at the position of the atoms using Rydberg-EIT spectroscopy of the microwave-induced avoided crossings. We produce $\sigma_-$, $\pi$, and $\sigma_+$ polarized microwaves with $>99$~\% fidelity and generate their combinations. We extend our purification techniques to frequencies away from Rydberg resonances by utilizing an auxiliary microwave field, generating two-photon microwave resonances. The techniques developed here will facilitate the engineering of dipolar interactions in atomic and molecular physics experiments. 
\end{abstract}


\maketitle

\section{Introduction}
Microwaves are a useful tool across quantum physics research. 
Precise control of the polarization of microwaves is especially important for a number of atomic and molecular physics platforms.
In the generation of circular Rydberg states, the purity of the microwave polarization is an important factor determining the state preparation fidelity~\cite{Signoles.2017, Cohen.2021}. 
Similarly, high-purity circularly polarized microwaves are needed to shield polar molecules from trap loss induced by inelastic collisions~\cite{Karman.2019,Anderegg.2021}. 
Microwave modification of the scattering properties of dipolar molecules enables evaporative cooling of molecules to degeneracy \cite{Valtolina.2020,Schindewolf.2022,Bigagli.2024}.
In these experiments, the microwaves couple rotational states of opposite parity, tuning the form and strength of the dipole-dipole interactions to generate a repulsive barrier, which modifies the scattering properties of the molecules \cite{Gorshkov.2008,Karman.2018,Karman.2025}. 

A similar approach can be used to tune the strength, angular dependence, and scaling of the interaction of a pair of Rydberg atoms. 
Enhancement of the interaction strength by resonant coupling of opposite-parity Rydberg states has been demonstrated in several experiments \cite{Tanasittikosol.2011,Brekke.2012,Xu.2024,Kurdak.2025}. 
Proposals to further engineer interactions, such as generating bound states, nullifying interactions, and realizing asymmetric blockade, require further control over the polarization of the microwave fields \cite{Sevinçli.2014, Shi.2017, Young.2021}.

In cold-atom experiments, reflections from nearby conducting surfaces can distort the polarization and field mode of the microwaves. 
When a specific microwave polarization is required for an experiment, the chamber and microwave sources are typically designed to mitigate this problem, often using glass cells and employing compensating sources with phase and amplitude control to empirically fine-tune the final polarization state \cite{Yuan.2023,Anderegg.2021,Schindewolf.2022,Bigagli.2023,Lin.2023}.

Here we describe a systematic approach to accurately create microwave fields of pure polarization in a steel vacuum chamber not designed for microwave control.
Our quantum optics experiment hosts a Rydberg-coupled cold-atom ensemble, which we use to probe the electric fields and their polarizations.
We apply microwaves to three sets of in-vacuum electrodes (originally designed to produce DC electric fields), with independent phase and amplitude control on each set. 
The fields produced by the three sources are characterized with avoided-crossing spectroscopy of Rydberg transitions.
We present our field characterization procedure and the construction of a ``chamber matrix'' that describes the linear mapping from the microwave control settings to the resulting electric field.
With knowledge of the mapping, we realize several different pre-defined polarizations with intensity fidelities $>99$~\%.
We also discuss the purification of microwave polarization away from resonance using an auxiliary field on two-photon resonance.

\section{Microwave Generation}

Our laser-cooled ensemble is surrounded by eight in-vacuum electrodes that control the DC and AC electric fields at the position of the atoms (see Fig.~\ref{fig:Fig1}). 
The microwave fields in the chamber are produced by driving three independent subsets of the electrodes (Fig.~\ref{fig:Fig1}(a)) chosen to avoid linear dependence between the electric fields generated by each set.
Shown in Fig.~\ref{fig:Fig1}(b) are the microwave components required for full control over the field generated by the source electrodes in our system.
We split the output of a single microwave generator into three branches, which are connected to the three sets of electrodes, providing stability of the relative phases. 
We control the amplitudes and relative phases of the sources with a three-channel electronically controlled attenuator and a pair of programmable phase shifters.  
Because the electrodes were designed to operate at DC, the D-Sub 9 feedthroughs are not 50 $\Omega$ impedance matched.
Monitoring the reflections from the chamber with a directional coupler, we use a stub tuner to reduce reflections and ensure a flat spectral response across an approximately $50$~MHz range centered on the desired frequency, which is sufficient to measure the polarization of the fields near the atomic resonance. 
We also use amplifiers and isolators in each branch (not shown in Fig. \ref{fig:Fig1}), to increase the input power and protect the various components. 

\section{Polarization Measurement}
The unknown polarizations of the electric fields from the three sources depend sensitively on the geometry of the electrodes and the surrounding environment.
The 5~cm to 10~cm microwave wavelengths in our experiment are similar to the length scales of our chamber, which encloses a variety of conducting surfaces.
Modeling this ``electrically challenging'' environment is difficult to perform and verify.
We therefore directly measure the fields using the ensemble of $^{87}$Rb atoms.

In recent years, Rydberg-based measurement of the polarization of microwave fields has been demonstrated in many experiments. 
Such experiments, performed with spin-unpolarized ensembles, measure only linearly polarized microwaves \cite{Sedlacek.2013,Cloutman.2024,You.2024,Cloutman.2025,Ren.2022,Bai.2019} or require a well-defined reference polarization \cite{Elgee.2024,Wang.2023,Yin.2024}.
Neither approach is suitable for generating and measuring arbitrary polarizations in an electrically challenging environment.

\begin{figure}[t!]
\centering
\includegraphics[width=\columnwidth]{"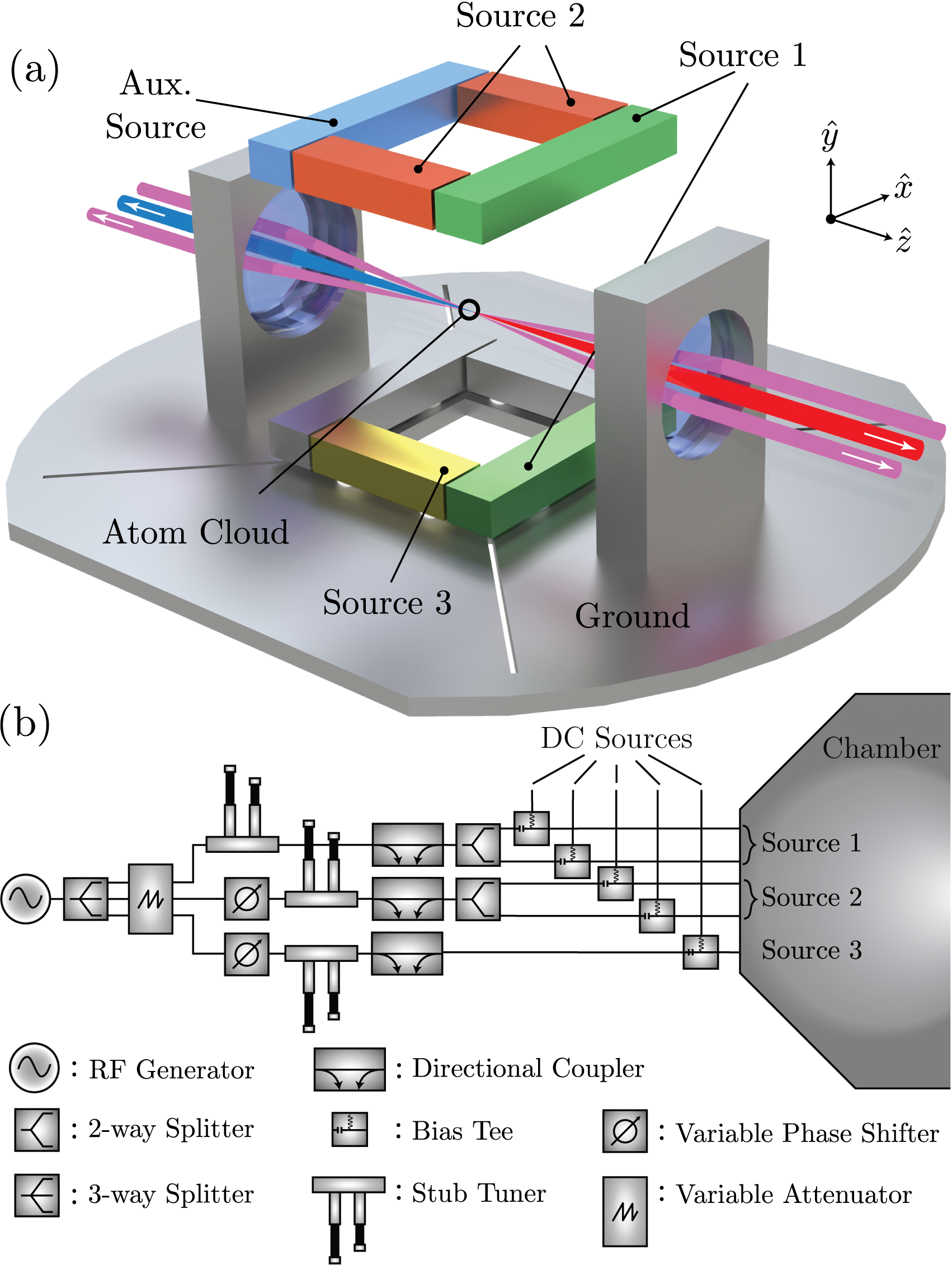"}
\caption{\label{fig:Fig1} (a) Two in-vacuum lenses focus the counterpropagating probe (red) and control (blue) laser beams onto the cold-atom ensemble, trapped in an optical dipole trap formed by a pair of 1012 nm beams (purple).
The eight electrodes were designed to cancel stray DC electric fields. Three sets of these electrodes (green, orange, and yellow) are used as primary microwave sources addressing the $88\,^{2}S_{1/2} -88\,^{2}P_{3/2}$ transition. An auxiliary source (blue electrode) addresses the $88\,^{2}P-87\,^{2}D$ transitions. (b) The output of a single signal generator is split into three branches, where the amplitude of each branch is controlled by a variable attenuator and the relative phases are tuned by a pair of variable phase shifters. Each source is impedance matched into the chamber using stub tuners. The DC field controls are combined with the microwaves using bias tees.}
\end{figure}

\begin{figure}[t!]
\centering
\includegraphics[width=\columnwidth]{"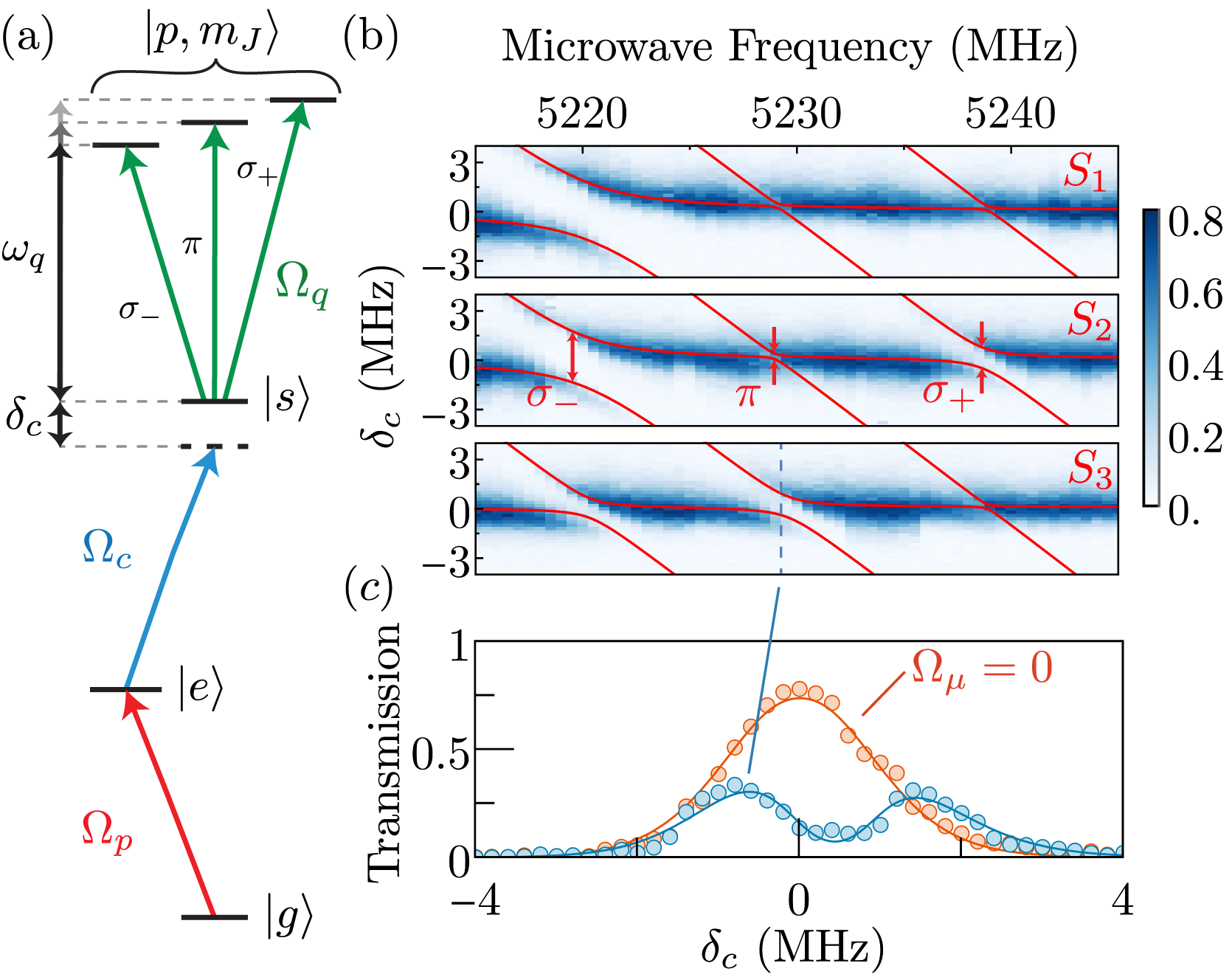"}
\caption{\label{fig:Fig2} (a) The probe and control fields couple the ensemble from the ground state $\ket{g}$ to the Rydberg state $\ket{s}$ through the intermediate state $\ket{e}$ with two-photon detuning $\delta_c$. The microwave fields, with polarization components $\sigma_-$, $\pi$, or $\sigma_+$, connect the Rydberg states $\ket{s}$ and $\ket{p,m_J}$ with resonant frequencies $\omega_{q}$ and Rabi frequencies $\Omega_{q}$. (b) The polarizations of the three sources $S_{1,2,3}$ are measured by scanning the microwave frequencies across the three resonances and observing their effect on the EIT transmission spectrum. The EIT scans were made by varying $\delta_c$ while keeping the probe fixed on resonance. 
Each plot corresponds to EIT spectroscopy with the field from a single source, where the color indicates the probe transmission through the cloud. (c) An EIT window unperturbed by microwaves (orange), and an Autler-Townes-split EIT feature (blue) sampled from the $\pi$ avoided crossing induced by $S_3$.}
\end{figure}

To characterize the polarization of the microwave fields, we perform electromagnetically induced transparency (EIT) \cite{Fleischhauer.2005} spectroscopy of Rydberg levels in a spin-polarized cloud of cold atoms. 
The ensemble is laser cooled to $\approx 10 \, \mu$K, loaded into a crossed optical dipole trap formed by a pair of 1012 nm beams (shown in purple in Fig.~\ref{fig:Fig1}(a)), and optically pumped to the $\ket{g}\equiv \ket{5\,^{2}S_{1/2},F=2, m_F = -2}$ state.
A weak probe laser, shown in red in Fig.~\ref{fig:Fig1}(a), is tuned to the $\ket{g}$ to $\ket{e}\equiv \ket{5\,^{2}P_{3/2},F^\prime = 3, m_F^\prime = -3}$ resonance and the transmission through the optically thick medium (optical depth = $20$-$30$) is measured.
A control field, shown in blue in Fig.~\ref{fig:Fig1}(a), connects the intermediate state $\ket{e}$ to the Rydberg state $\ket{s}\equiv\ket{88\,^{2}S_{1/2},m_J = -1/2}$, with a typical Rabi frequency of $\Omega_c/(2 \pi) \approx$ 6~MHz - 9~MHz. 
The strong control coupling opens an EIT window at two-photon resonance \cite{Fleischhauer.2005}. The level scheme is shown in Fig.~\ref{fig:Fig2}(a).
We cancel stray DC electric fields by applying static voltages to the eight in-vacuum electrodes.

The microwave fields address the $\ket{s}$ to $\ket{p,{m_J}}\equiv \ket{88\,^{2}P_{3/2},m_J}$ transitions, perturbing the Rydberg states and the position of the Rydberg-EIT window.
We apply a 0.5~mT magnetic field along the probe direction to lift the degeneracy of the $\ket{s}$ to $\ket{p,{m_J}}$ transitions coupled by the three possible microwave polarizations $\sigma_-$, $\pi$, and $\sigma_+$. 
(Unless otherwise noted, these polarizations are defined via a quantization axis pointing along the probe direction $\hat{z}$, as defined in Fig.~\ref{fig:Fig1}(a).) 
Each $\ket{s} \leftrightarrow\ket{p,m_J}$ transition gives rise to a spectrally resolved avoided crossing, which we measure by varying the control and microwave frequencies across the three resonances.
We show three examples of avoided crossing spectra in Fig.~\ref{fig:Fig2}(b), where we turn on each source one at a time.
Note that for all sources, the avoided crossings for all three polarizations have non-zero splittings. 
We show sample EIT transmission spectra in Fig.~\ref{fig:Fig2}(c). With no microwaves, a single EIT peak is present (red); when the field is applied on resonance, the bare state splits into a pair of dressed eigenstates whose splitting is related to the Rabi frequency of the microwave field.
These Rabi frequencies are extracted from the transmission spectra using a model of the probe susceptibility that accounts for the presence of control and microwave fields (red and blue solid lines).
The extracted Rabi frequencies are directly related to the field magnitudes $\mathcal{E}_{q} = \left|\hbar\Omega_{q}/d_q\right|$, where $d_q$ and $\Omega_{q}$ correspond to the dipole matrix element and Rabi frequency of polarization $q$.
The inferred electric fields are then provided as an input to a separate model that includes all $m_J$ states of the $\ket{s}$ and $\ket{p}$ manifolds to compute the eigenenergies, shown as the red lines overlaid on the data in Fig.~\ref{fig:Fig2}(b).

\section{Chamber Matrix}
Given three sources whose fields are linearly independent, an arbitrary polarization can be generated at the position of the atoms by adding the fields with the correct relative magnitudes and phases.
In our complicated microwave-reflecting environment, we must measure the fields produced by each source to determine the linear mapping between source control settings and generated fields.

We describe each source's control parameters (the field magnitude and phase) by the complex amplitude $V_\alpha = \mathcal{V}_\alpha e^{i \theta_\alpha}$, and we write the control parameters for all three sources in a single ``voltage vector''
\begin{equation}
 \mathbf{V} = \sum_\alpha V_\alpha \hat{\mathbf{v}}_{\alpha},
\end{equation}
where the basis vector $\hat{\mathbf{v}}_\alpha$  corresponds to only source $S_\alpha$ being on, e.g. $\hat{\mathbf{v}}_2 \equiv \left(0,1,0\right)$ corresponds to $S_2$ on and $S_1$, $S_3$  off.
In the abstract ``source" space, the $\left\{\hat{\mathbf{v}}_{\alpha}\right\}$ are orthonormal, $\hat{\mathbf{v}}_{\alpha} \cdot \hat{\mathbf{v}}_{\alpha^\prime} = \delta_{\alpha \alpha^\prime}$.

We define a ``chamber matrix'' $\mathbb{C}$, which relates the electric field $\mathbf{E}$ at the position of the atoms to $\mathbf{V}$ through $\mathbf{E} = \mathbb{C} \, \mathbf{V}$. We express $\mathbf{E}$ in the spherical vector basis, 
\begin{equation}
    \mathbf{E}=\sum_q  E_q  \hat{\boldsymbol{\epsilon}}_{q},
\end{equation}
where $\left\{\hat{\boldsymbol \epsilon}_q\right\}$  ($q=-1,0,1$) are the spherical-basis unit vectors corresponding to ($\sigma_-$, $\pi$, $\sigma_+$) polarizations.  $E_q$ are the complex amplitudes of the polarization components, $E_q = \mathcal{E}_q e^{i \phi_q}$. 
The linear mapping $\mathbb{C} $ can be written in terms of matrix elements $C_{q \alpha}$, so that the polarization component $E_{q^\prime}$ resulting from an applied $\mathbf{V}$ is 
\begin{equation}
E_{q^\prime} = \sum_{ \alpha}  \ C_{q^\prime \alpha} V_{\alpha}.
\end{equation}
The nine complex matrix elements  $C_{q \alpha}= \mathcal{C}_{q \alpha}\ e^{i \phi_{q \alpha}}$, characterized by 18 real parameters $\{ \mathcal{C}_{q \alpha},\ \phi_{q \alpha}\}$, completely determine the electric fields produced by an arbitrary $\mathbf{V}$.

Determining the matrix elements $\left\{C_{q \alpha}\right\}$ involves measuring the field parameters $\left\{\mathcal{E}_q,\phi_q\right\}$ as a function of the control parameters $\left\{\mathcal{V}_\alpha,\theta_\alpha\right\}$.
We choose the arbitrary global phase such that $\theta_1=0$, leaving us with five control parameters ($\mathcal{V}_1$,~$\mathcal{V}_2$,~$\mathcal{V}_3$,~$\theta_2$,~$\theta_3$), which are controlled by the variable attenuators and phase shifters shown in Fig.~\ref{fig:Fig1}(b).

\section{Chamber Matrix Characterization}
To measure the amplitudes $\mathcal{C}_{q\alpha}$ of the chamber matrix, we turn on one source at a time, as shown in Fig.~\ref{fig:Fig2}(b).
A given $\mathcal{C}_{q\alpha}$ determines the field amplitude of polarization $q$ generated by source $\alpha$, so the parameters $\{\mathcal{C}_{q\alpha}\}$ can be measured by relating the sizes of the avoided crossings to the field amplitudes, as described previously. 

\begin{figure}[t!]
\centering
\includegraphics[width=\columnwidth]{"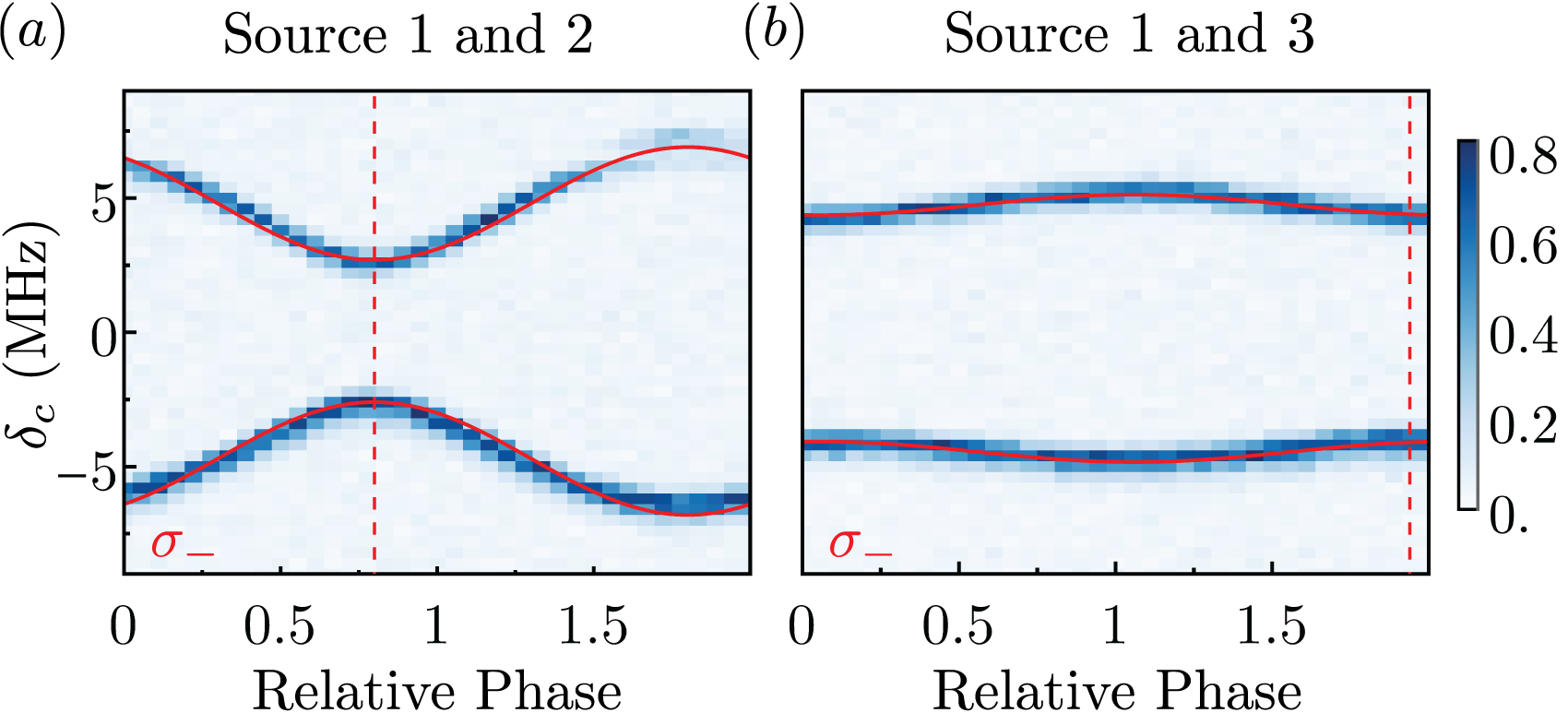"}
\caption{\label{fig:Fig3} Variation of the $q=-1$ avoided crossing size as a function of the relative phase (in multiples of $\pi$) between (a) sources 1 and 2 and (b) sources 1 and 3. The color indicates the probe transmission through the cloud. The solid red lines are sinusoidal fits to the EIT peak positions, and the dashed lines indicate the angle of maximum destructive interference. The fits are used to determine the phase difference between the same polarization component of two different sources. }
\end{figure}

The differences between the nine phases $\{\phi_{q\alpha}\}$ can be measured by interfering sources. 
Since our system is insensitive to changes in the global phase of the microwaves, we do not measure the absolute values of the parameters $\{\phi_{q\alpha}\}$, only the differences between them.
Turning on sources $\alpha$ and $\beta$ simultaneously ($\mathbf{V} = V_\alpha \hat{\mathbf{v}}_\alpha + V_\beta\hat{\mathbf{v}}_\beta$) generates a total field whose polarization component $q$ is given by, 
\begin{equation}
    E_{q} = \mathcal{C}_{q \alpha} \mathcal{V}_{\alpha} e^{i \left( \phi_{q \alpha} + \theta_\alpha \right) } + \mathcal{C}_{q \beta}  \mathcal{V}_{\beta} e^{i \left(\phi_{q \beta} + \theta_\beta \right) }.
\end{equation}
By varying one phase shifter angle and measuring $\mathcal{E}_{q}$, we map out the interference between sources $\alpha$ and $\beta$, and determine the difference between $\phi_{q\alpha}$ and $\phi_{q\beta}$ for each $q$.
We show two examples in Fig.~\ref{fig:Fig3}, where the amplitude of the total $\sigma_-$  field is measured as a function of phase difference between sources 1 and 2, and sources 1 and 3. 

For a given $q$, repeating this process for all pairs of sources yields the differences between the phase parameters $\phi_{q\alpha}$.
The phase difference measurement is repeated for all polarizations. 
At the magnetic fields we use, which clearly separate the three resonances (the Zeeman splittings are larger than the microwave Rabi frequencies), interference effects between fields with different $q$ are negligible. As a result, our measurements do not yield information about the phase relationships between orthogonal polarizations. We will discuss the consequences of this phase ambiguity below and present additional techniques to resolve it for applications where such phases are important

Having determined the relative phases between sources for a single polarization $q$, we proceed to find the voltage vector $\mathbf{V}^{(q)}$ that generates this polarization. 
Calculating $\mathbb{C}^{-1} \,  \hat{\boldsymbol{\epsilon}}_q$ results in an estimate of $\mathbf{V}^{(q)}$ which produces a field close to the desired polarization.
We then manually explore the nearby control parameter space to minimize the unwanted polarizations, improving the estimate of the voltage vector $\mathbf{V}^{(q)}$ and the corresponding polarization purity.
(Note that since the control parameter space is large, purifying the polarizations without the initial estimate $\mathbb{C}^{-1} \,  \hat{\boldsymbol{\epsilon}}_q$ would be challenging.)
\onecolumngrid

\begin{figure}[h] 
\includegraphics[width=\textwidth]{"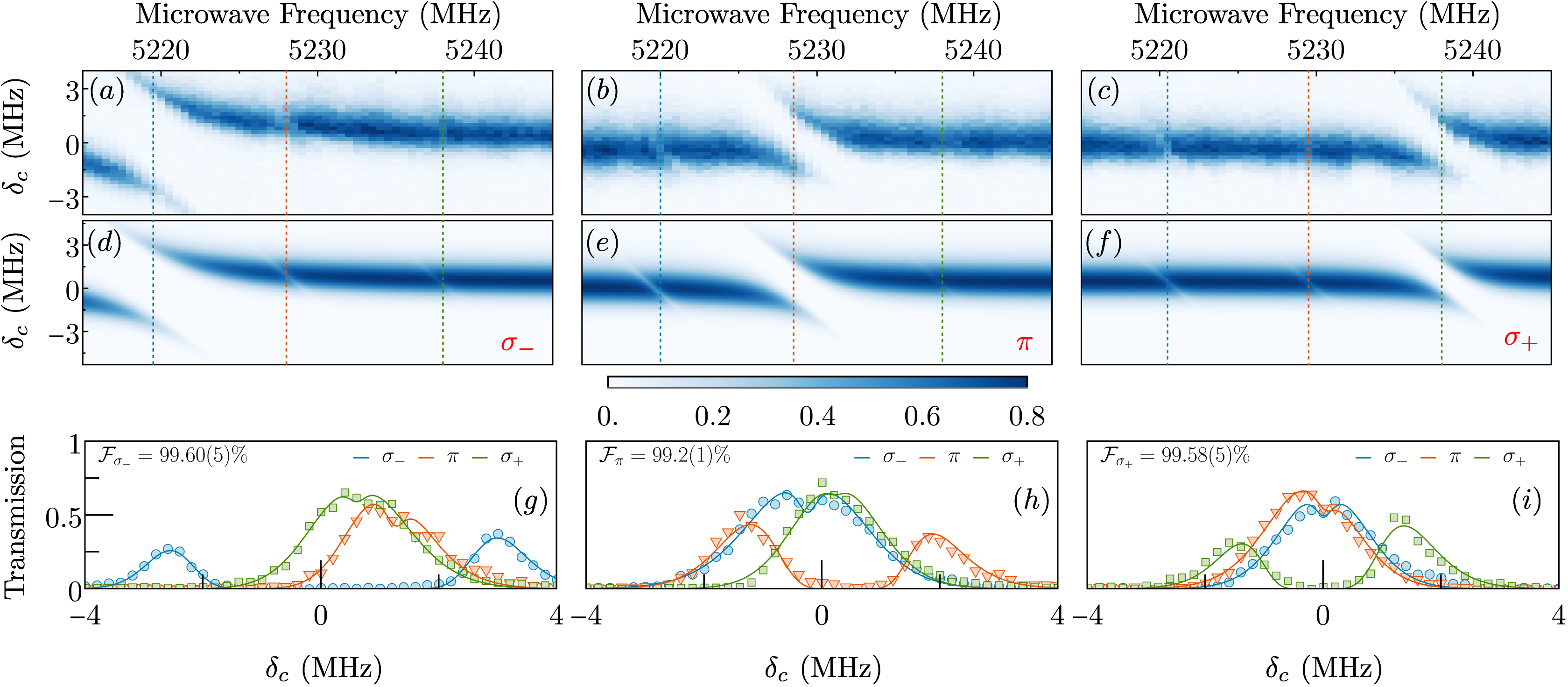"}
\caption{\label{fig:Fig4} Demonstration of high-purity microwave polarization. (a-c) Avoided-crossing spectra for the purified $\sigma_-$, $\pi$, and $\sigma_+$ polarizations. Avoided crossings from unwanted polarizations are nearly closed in all configurations. (d-f) Simulation of probe transmission using a model  including $\ket{g}$, $\ket{e}$, and all $m_J,m_I$ states of the Rydberg manifold. The model uses amplitudes extracted from the data in (g-i) using a four-level EIT model. (g-i) EIT signals at the avoided crossings (indicated by the dashed lines in (a-f)) for the $\sigma_-$, $\pi$, and $\sigma_+$  microwaves. The polarization amplitudes are extracted from the fits (solid lines). Our purification yields excellent fidelities, reaching $\mathcal{F}=$ 99.60(5)~\%, 99.2(1)~\%, and 99.58(5)~\% for the $\sigma_-$, $\pi$, and $\sigma_+$ calibrations, respectively.  The numbers in parenthesis indicate the $\pm1 \sigma$ statistical uncertainties. The corresponding normalized ratios of the field amplitudes are $(1 : 0.062 : 0.064)$, $(0.097 : 1 : 0.076)$, and $(0.063 : 0.066 : 1)$ for the three polarizations.}
\end{figure} 
\vspace{10pt}
\twocolumngrid
With manual improvement we obtain high-purity microwave fields $\mathbf{E}^{(q)}$ for each $q$, whose avoided-crossing spectra are shown in Fig.~\ref{fig:Fig4}(a-c).
These spectra indicate that for each polarization the undesirable avoided crossings have been closed. 
We define the fidelity, $\mathcal{F}_q $, of the electric field as $\mathcal{F}_q = \mathbf{E}^{(q)} \cdot \hat{\boldsymbol{\epsilon}}_q/\left| \mathbf{E}^{(q)}\right|$, and  find $\mathcal{F}_q>99\%$ for all three polarizations.
We also show in Fig.~\ref{fig:Fig4}(d-f) results from a probe susceptibility model that includes $\ket{g}$, $\ket{e}$, and all $m_J,m_I$ levels of the $88\,^{2}S_{1/2}$ and $88\,^{2}P_{3/2}$ manifolds. Fig.~\ref{fig:Fig4}(g-i) show corresponding transmission spectra at the positions of the avoided crossings. 

\begin{figure}[t]
\centering
\includegraphics[width=\columnwidth]{"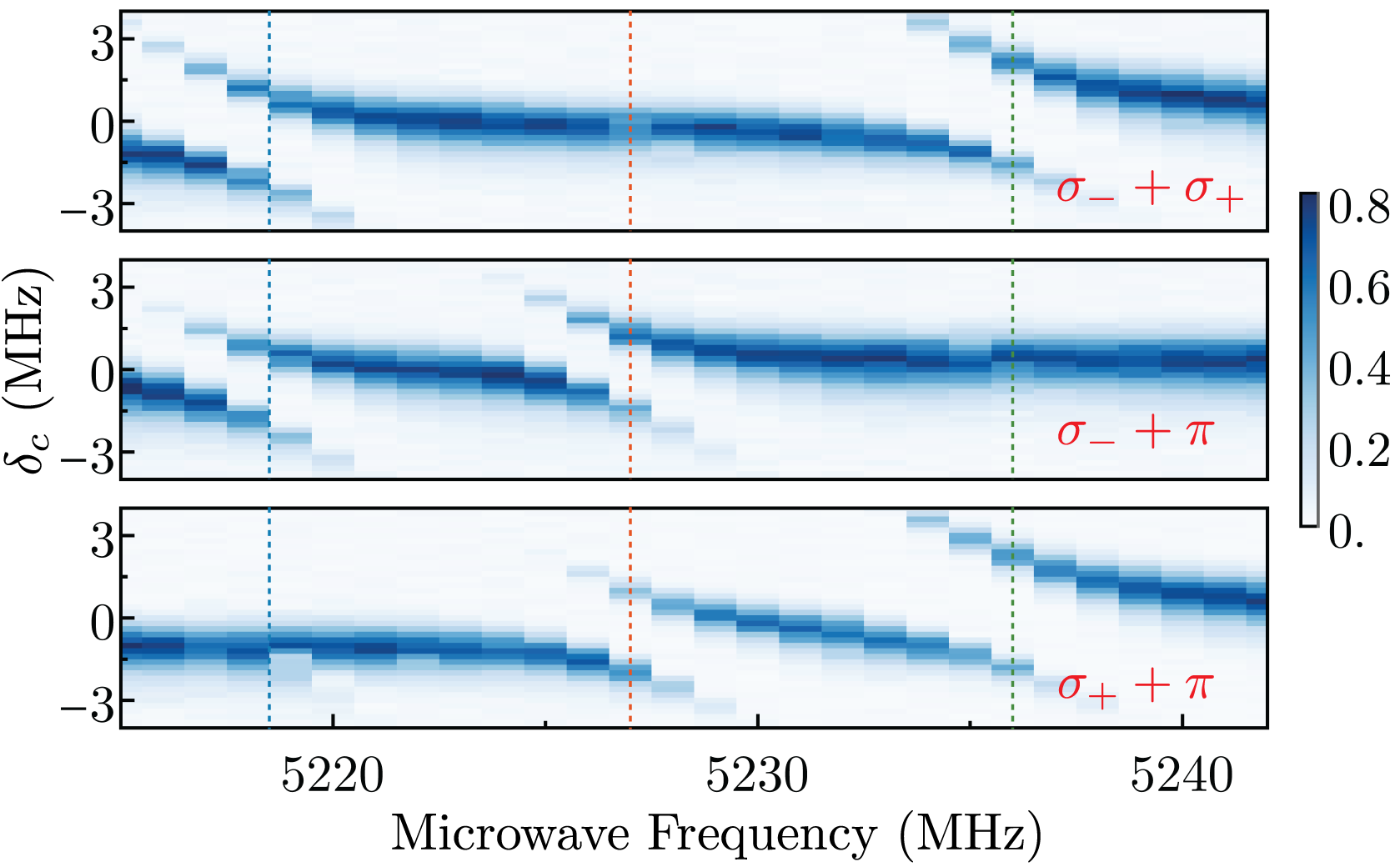"}
\caption{\label{fig:Fig5} Demonstration of superpositions of polarizations, achieved by adding the control settings used to generate each of the three orthogonal polarizations.}
\end{figure}

Since the control system is linear, combinations of polarizations can be constructed through linear combinations of the basis voltage vectors $\mathbf{V}^{(q)}$.
For example, the addition of the voltage vectors $\mathbf{V}^{\small{(+1)}}$ and $\mathbf{V}^{\small{(-1)}}$ generates a field with components along both $\hat{\boldsymbol{\epsilon}}_{\small{+1}}$ and $\hat{\boldsymbol{\epsilon}}_{\small-1}$ as shown in the top row of Fig.~\ref{fig:Fig5}.
As noted earlier, the measurement described above is insensitive to the phase differences between orthogonal polarizations. 
Thus, while we can precisely specify the desired intensities of each polarization, knowledge of the relative phase is required to determine the orientation of the total field. 
For example, if we generate a field with equal amplitudes of $\sigma_-$ and $\sigma_+$ via the control setting $\mathbf{V}^{{\small{(+1)}}} + \mathbf{V}^{{\small{(-1)}}}$ (with zero amplitude $\pi$), the resulting electric field is linearly polarized in the $xy$-plane. Since we did not measure the relative phase between  polarization components, the exact orientation is not known.

Determining the relative phase between polarization components can be performed through measurements that interfere components of the field with different polarizations,
 either by working at low magnetic field or by reorienting the measurement basis such that the new basis vectors $\hat{\epsilon}^\prime_q$ are linear combinations of the original $\hat{\epsilon}_q$, for example by rotating the magnetic field (and the optically pumped atomic states) to point along $\hat{x}$. 
We note, however, that many proposed microwave dressing schemes do not require knowledge of these phases \cite{Sevinçli.2014, Shi.2017, Young.2021}. 

\section{Off-Resonance Purification}
The techniques presented thus far achieve excellent polarization purity. 
However, in our system, the purity is limited in bandwidth: due to frequency-dependent impedance variations of our chamber, the fidelities we achieve are limited to a range of microwave frequencies $\approx 50$~MHz around the $88\,^{2}S_{1/2} \leftrightarrow 88\,^{2}P_{3/2}$ resonance. 
Generating pure polarization away from resonance is a key requirement for a number of experimental proposals with microwave-coupled Rydberg atoms \cite{Sevinçli.2014, Shi.2016,Shi.2017, Young.2021}.

\begin{figure}[b!]
\centering
\includegraphics[width=\columnwidth]{"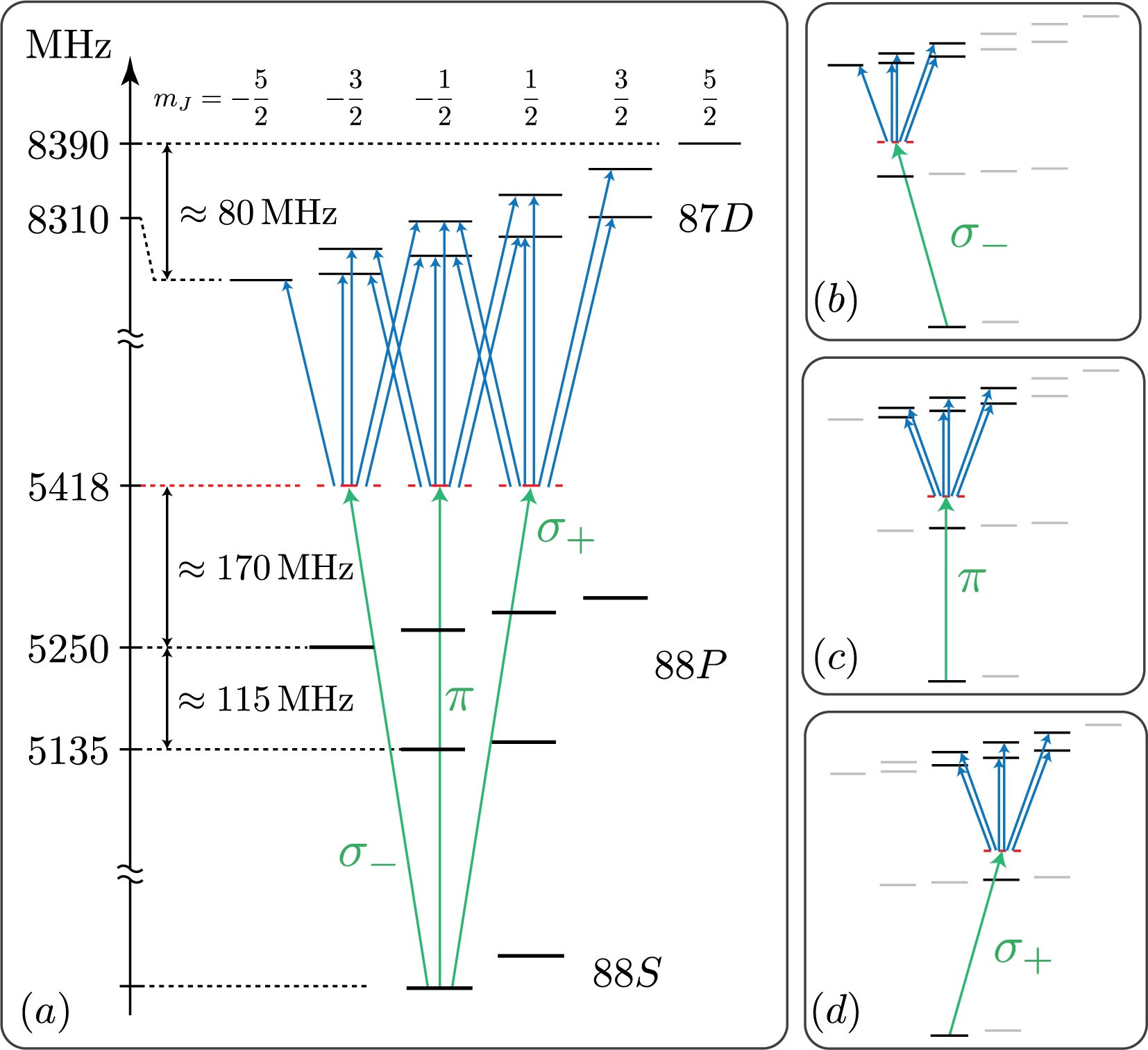"}
\caption{\label{fig:Fig6} Level scheme for off-resonance purification. (a) The primary microwave field (green) is blue detuned from the $88\,^{2}S_{1/2}-88\,^{2}P$ transition; with the introduction of an auxiliary microwave field (blue), resonant two-photon coupling is established between the $88\,^{2}S_{1/2}-87\,^{2}D$ states. The auxiliary field, composed of all three polarizations, is used to probe the components of the primary field, resonantly coupling $\ket{s}$ to nine states. (b-d) When the primary polarization is pure $\sigma_-$, $\pi$ or $\sigma_+$, only subsets of the avoided crossings are present. The other avoided crossings, corresponding to undesirable primary polarizations, are closed by tuning the control parameters.
}
\end{figure}

\begin{figure*}
\centering
\includegraphics[width=\textwidth]{"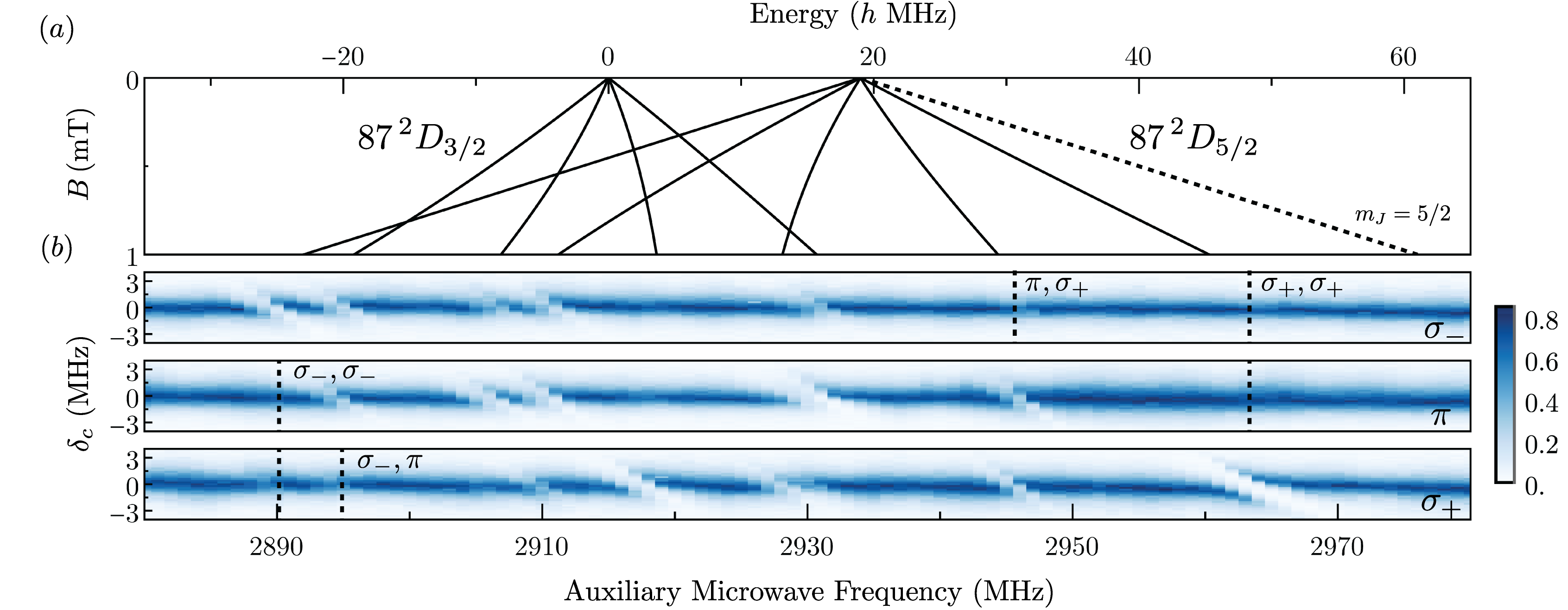"}
\caption{\label{fig:Fig7}  (a) At zero field, the $87\,^{2}D_{3/2}$ and $87\,^{2}D_{5/2}$ manifolds are split by $\approx 17$~MHz. Increasing the magnetic field strength lifts the degeneracy of the $m_J$ states and admixes states of definite $J$ into new eigenstates, with definite $m_J$, indicated by the solid lines. (b) Avoided crossing spectra as the auxiliary frequency is varied, with the primary microwaves fixed at 5418~MHz. The three plots show the spectra for the primary microwaves purified to the $\sigma_-$, $\pi$, and $\sigma_+$ polarizations, respectively. The absence of the two leftmost or rightmost avoided crossings is indicated by the vertical dashed lines, whose labels indicate the primary and auxiliary polarizations needed to open each crossing, absent in the purified case.}
\end{figure*}

To perform polarization measurements away from resonance, we introduce an auxiliary microwave field (using the blue electrode shown in Fig.~\ref{fig:Fig1}(a)) to create two-photon resonances (similar to Ref.~\cite{Liu.20228js}) between $\ket{s}$ and the $\ket{d, m_J} \equiv \ket{87\,^{2}D, m_J}$ manifold.
The level scheme is shown in Fig.~\ref{fig:Fig6}(a).
The primary field (shown in green), whose polarization we want to measure, may have nonzero amplitudes in all three polarizations prior to purification.
Applying the auxiliary field (shown in blue) provides resonant two-photon coupling to a total of nine states in the $\ket{d}$ manifold.
To spectrally resolve the resonances, we apply a $1$~mT magnetic field, splitting the $m_J$ states across $80$ MHz. 
(Note that the fine structure splitting of the $87\,^{2}D$ states is small compared to the total Zeeman shift, so the $^{2}D_{3/2}$ and $^{2}D_{5/2}$ manifolds are mixed.)
Then, as above, we perform avoided-crossing spectroscopy by scanning the auxiliary microwave frequency over the two-photon resonances.

The size of each avoided crossing is determined by the complex sum of the Rabi frequencies of each two-photon transition from $\ket{s}$ to a given $\ket{d, m_J}$.
Due to interference between these pathways, it is challenging to generalize the techniques of previous sections when the auxiliary microwave polarization is not known \textit{a priori}.
Instead, to simplify the measurement, we use only the transitions from $\ket{s}$ to $\ket{d, -5/2}$ and to $\ket{d, +3/2}$, which have only a single excitation pathway. (This is possible with a $d$ manifold but not with an $s$ manifold).
Using these states as resources, as shown in Fig.~\ref{fig:Fig6}(b-d), we can eliminate undesirable polarizations.
To achieve pure $\pi$ polarization for the primary microwave field, the phase and amplitude controls over the three sources can be varied until the $m_J = -5/2$ and $+3/2$ avoided crossings simultaneously close.
Similarly, to purify the polarization to $\sigma_+$ or $\sigma_-$, the avoided crossings corresponding to $m_J= -5/2, \, -3/2$ and $m_J= +1/2, \, +3/2$ can be closed, respectively. 
The only requirement is that all polarizations of the auxiliary microwave field should be nonzero across the frequency range of interest.

We perform off-resonant purification with the primary field frequency set to 5418~MHz, $\approx 200$ MHz detuned from the unperturbed $88\,^{2}S_{1/2}-88\,^{2}P_{3/2}$ resonance (and well outside our typical purification bandwidth).
The auxiliary microwave frequency is varied to span the nine distinct two-photon resonances shown in Fig.~\ref{fig:Fig7}(a).
Shown in Fig.~\ref{fig:Fig7}(b) are three identical auxiliary frequency scans, where only the primary microwave polarizations were changed in each plot.
To purify the microwaves to $\sigma_-$, the avoided crossings at 2947~MHz ($\ket{s}-\ket{d,+1/2}$) and 2965~MHz ($\ket{s}-\ket{d,+3/2}$) were closed by manually searching the control parameter space (top panel of Fig.~\ref{fig:Fig7}(b)). 
The primary microwaves were purified to $\pi$ and $\sigma_+$ in a similar manner (lower two panels of Fig.~\ref{fig:Fig7}(b)).
In the same fashion as Fig.~\ref{fig:Fig5}, the voltage settings yielding the orthogonal polarizations can be combined to form arbitrary polarization states. 

In principle, this approach can be extended farther off-resonance by increasing the primary and auxiliary powers to retain sufficiently large avoided crossings.
The primary challenge  is that accidental resonances may make certain regions of the parameter space inaccessible for a given choice of $S, \, P,$ and $D$ states.
Therefore, a careful survey of the energy landscape is necessary to find the two-photon resonances that provide the cleanest signals. 

\section{Conclusion}
In summary, we have presented techniques for the generation and characterization of high-purity microwave fields in an electrically challenging stainless steel vacuum chamber using resonant microwave coupling of Rydberg levels.
The ability to create well-defined fields in such a complex environment relaxes the constraints on apparatus design due to the need for polarization control, allowing more flexibility in building atomic physics experiments. 
Utilizing a two-photon process, we extended the purification technique away from Rydberg resonances.
Electronic control of the microwaves allows for dynamical variation of microwave polarizations as well as the rapid characterization of the dependence of Rydberg atomic properties on the polarization and detuning of microwave-dressing parameters.
Precise control over this challenging degree of freedom will expand the resources available to engineer the properties of atomic systems ranging from Rydberg atoms to ultracold dipolar molecules.

\textit{Acknowledgments }--- 
The authors thank Daniel S. Barker, Eric C. Benck, and Alan Migdall for their critical review of the manuscript. 
D.K., P.R.B., and Y.L. were supported by the MAQP grant W911NF2420107. P.R.B. ~also acknowledges the support of the ARCS Foundation, Metro-Washington Chapter.  

\textit{Data Availability }--- 
The data that support the findings of this paper are openly available \cite{Zenodo.2025}.



\begin{thebibliography}{33}%
\makeatletter
\providecommand \@ifxundefined [1]{%
 \@ifx{#1\undefined}
}%
\providecommand \@ifnum [1]{%
 \ifnum #1\expandafter \@firstoftwo
 \else \expandafter \@secondoftwo
 \fi
}%
\providecommand \@ifx [1]{%
 \ifx #1\expandafter \@firstoftwo
 \else \expandafter \@secondoftwo
 \fi
}%
\providecommand \natexlab [1]{#1}%
\providecommand \enquote  [1]{``#1''}%
\providecommand \bibnamefont  [1]{#1}%
\providecommand \bibfnamefont [1]{#1}%
\providecommand \citenamefont [1]{#1}%
\providecommand \href@noop [0]{\@secondoftwo}%
\providecommand \href [0]{\begingroup \@sanitize@url \@href}%
\providecommand \@href[1]{\@@startlink{#1}\@@href}%
\providecommand \@@href[1]{\endgroup#1\@@endlink}%
\providecommand \@sanitize@url [0]{\catcode `\\12\catcode `\$12\catcode
  `\&12\catcode `\#12\catcode `\^12\catcode `\_12\catcode `\%12\relax}%
\providecommand \@@startlink[1]{}%
\providecommand \@@endlink[0]{}%
\providecommand \url  [0]{\begingroup\@sanitize@url \@url }%
\providecommand \@url [1]{\endgroup\@href {#1}{\urlprefix }}%
\providecommand \urlprefix  [0]{URL }%
\providecommand \Eprint [0]{\href }%
\providecommand \doibase [0]{https://doi.org/}%
\providecommand \selectlanguage [0]{\@gobble}%
\providecommand \bibinfo  [0]{\@secondoftwo}%
\providecommand \bibfield  [0]{\@secondoftwo}%
\providecommand \translation [1]{[#1]}%
\providecommand \BibitemOpen [0]{}%
\providecommand \bibitemStop [0]{}%
\providecommand \bibitemNoStop [0]{.\EOS\space}%
\providecommand \EOS [0]{\spacefactor3000\relax}%
\providecommand \BibitemShut  [1]{\csname bibitem#1\endcsname}%
\let\auto@bib@innerbib\@empty
\bibitem [{\citenamefont {Signoles}\ \emph {et~al.}(2017)\citenamefont
  {Signoles}, \citenamefont {Dietsche}, \citenamefont {Facon}, \citenamefont
  {Grosso}, \citenamefont {Haroche}, \citenamefont {Raimond}, \citenamefont
  {Brune},\ and\ \citenamefont {Gleyzes}}]{Signoles.2017}%
  \BibitemOpen
  \bibfield  {author} {\bibinfo {author} {\bibfnamefont {A.}~\bibnamefont
  {Signoles}}, \bibinfo {author} {\bibfnamefont {E.~K.}\ \bibnamefont
  {Dietsche}}, \bibinfo {author} {\bibfnamefont {A.}~\bibnamefont {Facon}},
  \bibinfo {author} {\bibfnamefont {D.}~\bibnamefont {Grosso}}, \bibinfo
  {author} {\bibfnamefont {S.}~\bibnamefont {Haroche}}, \bibinfo {author}
  {\bibfnamefont {J.~M.}\ \bibnamefont {Raimond}}, \bibinfo {author}
  {\bibfnamefont {M.}~\bibnamefont {Brune}},\ and\ \bibinfo {author}
  {\bibfnamefont {S.}~\bibnamefont {Gleyzes}},\ }\bibfield  {title} {\bibinfo
  {title} {{Coherent Transfer between Low-Angular-Momentum and Circular Rydberg
  States}},\ }\href {https://doi.org/10.1103/physrevlett.118.253603} {\bibfield
   {journal} {\bibinfo  {journal} {Physical Review Letters}\ }\textbf {\bibinfo
  {volume} {118}},\ \bibinfo {pages} {253603} (\bibinfo {year} {2017})},\
  \Eprint {https://arxiv.org/abs/1703.05918} {1703.05918} \BibitemShut
  {NoStop}%
\bibitem [{\citenamefont {Cohen}\ and\ \citenamefont
  {Thompson}(2021)}]{Cohen.2021}%
  \BibitemOpen
  \bibfield  {author} {\bibinfo {author} {\bibfnamefont {S.~R.}\ \bibnamefont
  {Cohen}}\ and\ \bibinfo {author} {\bibfnamefont {J.~D.}\ \bibnamefont
  {Thompson}},\ }\bibfield  {title} {\bibinfo {title} {{Quantum Computing with
  Circular Rydberg Atoms}},\ }\href
  {https://doi.org/10.1103/prxquantum.2.030322} {\bibfield  {journal} {\bibinfo
   {journal} {PRX Quantum}\ }\textbf {\bibinfo {volume} {2}},\ \bibinfo {pages}
  {030322} (\bibinfo {year} {2021})},\ \Eprint
  {https://arxiv.org/abs/2103.12744} {2103.12744} \BibitemShut {NoStop}%
\bibitem [{\citenamefont {Karman}\ and\ \citenamefont
  {Hutson}(2019)}]{Karman.2019}%
  \BibitemOpen
  \bibfield  {author} {\bibinfo {author} {\bibfnamefont {T.}~\bibnamefont
  {Karman}}\ and\ \bibinfo {author} {\bibfnamefont {J.~M.}\ \bibnamefont
  {Hutson}},\ }\bibfield  {title} {\bibinfo {title} {{Microwave shielding of
  ultracold polar molecules with imperfectly circular polarization}},\ }\href
  {https://doi.org/10.1103/physreva.100.052704} {\bibfield  {journal} {\bibinfo
   {journal} {Physical Review A}\ }\textbf {\bibinfo {volume} {100}},\ \bibinfo
  {pages} {052704} (\bibinfo {year} {2019})},\ \Eprint
  {https://arxiv.org/abs/1908.01759} {1908.01759} \BibitemShut {NoStop}%
\bibitem [{\citenamefont {Anderegg}\ \emph {et~al.}(2021)\citenamefont
  {Anderegg}, \citenamefont {Burchesky}, \citenamefont {Bao}, \citenamefont
  {Yu}, \citenamefont {Karman}, \citenamefont {Chae}, \citenamefont {Ni},
  \citenamefont {Ketterle},\ and\ \citenamefont {Doyle}}]{Anderegg.2021}%
  \BibitemOpen
  \bibfield  {author} {\bibinfo {author} {\bibfnamefont {L.}~\bibnamefont
  {Anderegg}}, \bibinfo {author} {\bibfnamefont {S.}~\bibnamefont {Burchesky}},
  \bibinfo {author} {\bibfnamefont {Y.}~\bibnamefont {Bao}}, \bibinfo {author}
  {\bibfnamefont {S.~S.}\ \bibnamefont {Yu}}, \bibinfo {author} {\bibfnamefont
  {T.}~\bibnamefont {Karman}}, \bibinfo {author} {\bibfnamefont
  {E.}~\bibnamefont {Chae}}, \bibinfo {author} {\bibfnamefont {K.-K.}\
  \bibnamefont {Ni}}, \bibinfo {author} {\bibfnamefont {W.}~\bibnamefont
  {Ketterle}},\ and\ \bibinfo {author} {\bibfnamefont {J.~M.}\ \bibnamefont
  {Doyle}},\ }\bibfield  {title} {\bibinfo {title} {{Observation of microwave
  shielding of ultracold molecules}},\ }\href
  {https://doi.org/10.1126/science.abg9502} {\bibfield  {journal} {\bibinfo
  {journal} {Science}\ }\textbf {\bibinfo {volume} {373}},\ \bibinfo {pages}
  {779} (\bibinfo {year} {2021})},\ \Eprint {https://arxiv.org/abs/2102.04365}
  {2102.04365} \BibitemShut {NoStop}%
\bibitem [{\citenamefont {Valtolina}\ \emph {et~al.}(2020)\citenamefont
  {Valtolina}, \citenamefont {Matsuda}, \citenamefont {Tobias}, \citenamefont
  {Li}, \citenamefont {Marco},\ and\ \citenamefont {Ye}}]{Valtolina.2020}%
  \BibitemOpen
  \bibfield  {author} {\bibinfo {author} {\bibfnamefont {G.}~\bibnamefont
  {Valtolina}}, \bibinfo {author} {\bibfnamefont {K.}~\bibnamefont {Matsuda}},
  \bibinfo {author} {\bibfnamefont {W.~G.}\ \bibnamefont {Tobias}}, \bibinfo
  {author} {\bibfnamefont {J.-R.}\ \bibnamefont {Li}}, \bibinfo {author}
  {\bibfnamefont {L.~D.}\ \bibnamefont {Marco}},\ and\ \bibinfo {author}
  {\bibfnamefont {J.}~\bibnamefont {Ye}},\ }\bibfield  {title} {\bibinfo
  {title} {{Dipolar evaporation of reactive molecules to below the Fermi
  temperature}},\ }\href {https://doi.org/10.1038/s41586-020-2980-7} {\bibfield
   {journal} {\bibinfo  {journal} {Nature}\ }\textbf {\bibinfo {volume}
  {588}},\ \bibinfo {pages} {239} (\bibinfo {year} {2020})},\ \Eprint
  {https://arxiv.org/abs/2007.12277} {2007.12277} \BibitemShut {NoStop}%
\bibitem [{\citenamefont {Schindewolf}\ \emph {et~al.}(2022)\citenamefont
  {Schindewolf}, \citenamefont {Bause}, \citenamefont {Chen}, \citenamefont
  {Duda}, \citenamefont {Karman}, \citenamefont {Bloch},\ and\ \citenamefont
  {Luo}}]{Schindewolf.2022}%
  \BibitemOpen
  \bibfield  {author} {\bibinfo {author} {\bibfnamefont {A.}~\bibnamefont
  {Schindewolf}}, \bibinfo {author} {\bibfnamefont {R.}~\bibnamefont {Bause}},
  \bibinfo {author} {\bibfnamefont {X.-Y.}\ \bibnamefont {Chen}}, \bibinfo
  {author} {\bibfnamefont {M.}~\bibnamefont {Duda}}, \bibinfo {author}
  {\bibfnamefont {T.}~\bibnamefont {Karman}}, \bibinfo {author} {\bibfnamefont
  {I.}~\bibnamefont {Bloch}},\ and\ \bibinfo {author} {\bibfnamefont {X.-Y.}\
  \bibnamefont {Luo}},\ }\bibfield  {title} {\bibinfo {title} {{Evaporation of
  microwave-shielded polar molecules to quantum degeneracy}},\ }\href
  {https://doi.org/10.1038/s41586-022-04900-0} {\bibfield  {journal} {\bibinfo
  {journal} {Nature}\ }\textbf {\bibinfo {volume} {607}},\ \bibinfo {pages}
  {677} (\bibinfo {year} {2022})},\ \Eprint {https://arxiv.org/abs/2201.05143}
  {2201.05143} \BibitemShut {NoStop}%
\bibitem [{\citenamefont {Bigagli}\ \emph {et~al.}(2024)\citenamefont
  {Bigagli}, \citenamefont {Yuan}, \citenamefont {Zhang}, \citenamefont
  {Bulatovic}, \citenamefont {Karman}, \citenamefont {Stevenson},\ and\
  \citenamefont {Will}}]{Bigagli.2024}%
  \BibitemOpen
  \bibfield  {author} {\bibinfo {author} {\bibfnamefont {N.}~\bibnamefont
  {Bigagli}}, \bibinfo {author} {\bibfnamefont {W.}~\bibnamefont {Yuan}},
  \bibinfo {author} {\bibfnamefont {S.}~\bibnamefont {Zhang}}, \bibinfo
  {author} {\bibfnamefont {B.}~\bibnamefont {Bulatovic}}, \bibinfo {author}
  {\bibfnamefont {T.}~\bibnamefont {Karman}}, \bibinfo {author} {\bibfnamefont
  {I.}~\bibnamefont {Stevenson}},\ and\ \bibinfo {author} {\bibfnamefont
  {S.}~\bibnamefont {Will}},\ }\bibfield  {title} {\bibinfo {title}
  {{Observation of Bose–Einstein condensation of dipolar molecules}},\ }\href
  {https://doi.org/10.1038/s41586-024-07492-z} {\bibfield  {journal} {\bibinfo
  {journal} {Nature}\ }\textbf {\bibinfo {volume} {631}},\ \bibinfo {pages}
  {289} (\bibinfo {year} {2024})},\ \Eprint {https://arxiv.org/abs/2312.10965}
  {2312.10965} \BibitemShut {NoStop}%
\bibitem [{\citenamefont {Gorshkov}\ \emph {et~al.}(2008)\citenamefont
  {Gorshkov}, \citenamefont {Rabl}, \citenamefont {Pupillo}, \citenamefont
  {Micheli}, \citenamefont {Zoller}, \citenamefont {Lukin},\ and\ \citenamefont
  {Büchler}}]{Gorshkov.2008}%
  \BibitemOpen
  \bibfield  {author} {\bibinfo {author} {\bibfnamefont {A.~V.}\ \bibnamefont
  {Gorshkov}}, \bibinfo {author} {\bibfnamefont {P.}~\bibnamefont {Rabl}},
  \bibinfo {author} {\bibfnamefont {G.}~\bibnamefont {Pupillo}}, \bibinfo
  {author} {\bibfnamefont {A.}~\bibnamefont {Micheli}}, \bibinfo {author}
  {\bibfnamefont {P.}~\bibnamefont {Zoller}}, \bibinfo {author} {\bibfnamefont
  {M.~D.}\ \bibnamefont {Lukin}},\ and\ \bibinfo {author} {\bibfnamefont
  {H.~P.}\ \bibnamefont {Büchler}},\ }\bibfield  {title} {\bibinfo {title}
  {{Suppression of Inelastic Collisions Between Polar Molecules With a
  Repulsive Shield}},\ }\href {https://doi.org/10.1103/physrevlett.101.073201}
  {\bibfield  {journal} {\bibinfo  {journal} {Physical Review Letters}\
  }\textbf {\bibinfo {volume} {101}},\ \bibinfo {pages} {073201} (\bibinfo
  {year} {2008})},\ \Eprint {https://arxiv.org/abs/0805.0457} {0805.0457}
  \BibitemShut {NoStop}%
\bibitem [{\citenamefont {Karman}\ and\ \citenamefont
  {Hutson}(2018)}]{Karman.2018}%
  \BibitemOpen
  \bibfield  {author} {\bibinfo {author} {\bibfnamefont {T.}~\bibnamefont
  {Karman}}\ and\ \bibinfo {author} {\bibfnamefont {J.~M.}\ \bibnamefont
  {Hutson}},\ }\bibfield  {title} {\bibinfo {title} {{Microwave Shielding of
  Ultracold Polar Molecules}},\ }\href
  {https://doi.org/10.1103/physrevlett.121.163401} {\bibfield  {journal}
  {\bibinfo  {journal} {Physical Review Letters}\ }\textbf {\bibinfo {volume}
  {121}},\ \bibinfo {pages} {163401} (\bibinfo {year} {2018})},\ \Eprint
  {https://arxiv.org/abs/1806.03608} {1806.03608} \BibitemShut {NoStop}%
\bibitem [{\citenamefont {Karman}\ \emph {et~al.}(2025)\citenamefont {Karman},
  \citenamefont {Bigagli}, \citenamefont {Yuan}, \citenamefont {Zhang},
  \citenamefont {Stevenson},\ and\ \citenamefont {Will}}]{Karman.2025}%
  \BibitemOpen
  \bibfield  {author} {\bibinfo {author} {\bibfnamefont {T.}~\bibnamefont
  {Karman}}, \bibinfo {author} {\bibfnamefont {N.}~\bibnamefont {Bigagli}},
  \bibinfo {author} {\bibfnamefont {W.}~\bibnamefont {Yuan}}, \bibinfo {author}
  {\bibfnamefont {S.}~\bibnamefont {Zhang}}, \bibinfo {author} {\bibfnamefont
  {I.}~\bibnamefont {Stevenson}},\ and\ \bibinfo {author} {\bibfnamefont
  {S.}~\bibnamefont {Will}},\ }\bibfield  {title} {\bibinfo {title} {{Double
  Microwave Shielding}},\ }\bibfield  {journal} {\bibinfo  {journal} {arXiv}\
  }\href {https://doi.org/10.48550/arxiv.2501.08095}
  {10.48550/arxiv.2501.08095} (\bibinfo {year} {2025}),\ \Eprint
  {https://arxiv.org/abs/2501.08095} {2501.08095} \BibitemShut {NoStop}%
\bibitem [{\citenamefont {Tanasittikosol}\ \emph {et~al.}(2011)\citenamefont
  {Tanasittikosol}, \citenamefont {Pritchard}, \citenamefont {Maxwell},
  \citenamefont {Gauguet}, \citenamefont {Weatherill}, \citenamefont
  {Potvliege},\ and\ \citenamefont {Adams}}]{Tanasittikosol.2011}%
  \BibitemOpen
  \bibfield  {author} {\bibinfo {author} {\bibfnamefont {M.}~\bibnamefont
  {Tanasittikosol}}, \bibinfo {author} {\bibfnamefont {J.~D.}\ \bibnamefont
  {Pritchard}}, \bibinfo {author} {\bibfnamefont {D.}~\bibnamefont {Maxwell}},
  \bibinfo {author} {\bibfnamefont {A.}~\bibnamefont {Gauguet}}, \bibinfo
  {author} {\bibfnamefont {K.~J.}\ \bibnamefont {Weatherill}}, \bibinfo
  {author} {\bibfnamefont {R.~M.}\ \bibnamefont {Potvliege}},\ and\ \bibinfo
  {author} {\bibfnamefont {C.~S.}\ \bibnamefont {Adams}},\ }\bibfield  {title}
  {\bibinfo {title} {{Microwave dressing of Rydberg dark states}},\ }\href
  {https://doi.org/10.1088/0953-4075/44/18/184020} {\bibfield  {journal}
  {\bibinfo  {journal} {Journal of Physics B: Atomic, Molecular and Optical
  Physics}\ }\textbf {\bibinfo {volume} {44}},\ \bibinfo {pages} {184020}
  (\bibinfo {year} {2011})},\ \Eprint {https://arxiv.org/abs/1102.0226}
  {1102.0226} \BibitemShut {NoStop}%
\bibitem [{\citenamefont {Brekke}\ \emph {et~al.}(2012)\citenamefont {Brekke},
  \citenamefont {Day},\ and\ \citenamefont {Walker}}]{Brekke.2012}%
  \BibitemOpen
  \bibfield  {author} {\bibinfo {author} {\bibfnamefont {E.}~\bibnamefont
  {Brekke}}, \bibinfo {author} {\bibfnamefont {J.~O.}\ \bibnamefont {Day}},\
  and\ \bibinfo {author} {\bibfnamefont {T.~G.}\ \bibnamefont {Walker}},\
  }\bibfield  {title} {\bibinfo {title} {{Excitation suppression due to
  interactions between microwave-dressed Rydberg atoms}},\ }\href
  {https://doi.org/10.1103/physreva.86.033406} {\bibfield  {journal} {\bibinfo
  {journal} {Physical Review A}\ }\textbf {\bibinfo {volume} {86}},\ \bibinfo
  {pages} {033406} (\bibinfo {year} {2012})},\ \Eprint
  {https://arxiv.org/abs/1207.3006} {1207.3006} \BibitemShut {NoStop}%
\bibitem [{\citenamefont {Xu}\ \emph {et~al.}(2024)\citenamefont {Xu},
  \citenamefont {Ye}, \citenamefont {Chang}, \citenamefont {Shi},\ and\
  \citenamefont {Li}}]{Xu.2024}%
  \BibitemOpen
  \bibfield  {author} {\bibinfo {author} {\bibfnamefont {B.}~\bibnamefont
  {Xu}}, \bibinfo {author} {\bibfnamefont {G.-S.}\ \bibnamefont {Ye}}, \bibinfo
  {author} {\bibfnamefont {Y.}~\bibnamefont {Chang}}, \bibinfo {author}
  {\bibfnamefont {T.}~\bibnamefont {Shi}},\ and\ \bibinfo {author}
  {\bibfnamefont {L.}~\bibnamefont {Li}},\ }\bibfield  {title} {\bibinfo
  {title} {{Continuously tunable single-photon level nonlinearity with Rydberg
  state wave-function engineering}},\ }\href
  {https://doi.org/10.1088/1361-6633/ad847e} {\bibfield  {journal} {\bibinfo
  {journal} {Reports on Progress in Physics}\ }\textbf {\bibinfo {volume}
  {87}},\ \bibinfo {pages} {110502} (\bibinfo {year} {2024})}\BibitemShut
  {NoStop}%
\bibitem [{\citenamefont {Kurdak}\ \emph
  {et~al.}(2025{\natexlab{a}})\citenamefont {Kurdak}, \citenamefont {Banner},
  \citenamefont {Li}, \citenamefont {Muleady}, \citenamefont {Gorshkov},
  \citenamefont {Rolston},\ and\ \citenamefont {Porto}}]{Kurdak.2025}%
  \BibitemOpen
  \bibfield  {author} {\bibinfo {author} {\bibfnamefont {D.}~\bibnamefont
  {Kurdak}}, \bibinfo {author} {\bibfnamefont {P.~R.}\ \bibnamefont {Banner}},
  \bibinfo {author} {\bibfnamefont {Y.}~\bibnamefont {Li}}, \bibinfo {author}
  {\bibfnamefont {S.~R.}\ \bibnamefont {Muleady}}, \bibinfo {author}
  {\bibfnamefont {A.~V.}\ \bibnamefont {Gorshkov}}, \bibinfo {author}
  {\bibfnamefont {S.~L.}\ \bibnamefont {Rolston}},\ and\ \bibinfo {author}
  {\bibfnamefont {J.~V.}\ \bibnamefont {Porto}},\ }\bibfield  {title} {\bibinfo
  {title} {{Enhancement of Rydberg Blockade via Microwave Dressing}},\ }\href
  {https://doi.org/10.1103/physrevlett.134.123404} {\bibfield  {journal}
  {\bibinfo  {journal} {Physical Review Letters}\ }\textbf {\bibinfo {volume}
  {134}},\ \bibinfo {pages} {123404} (\bibinfo {year} {2025}{\natexlab{a}})},\
  \Eprint {https://arxiv.org/abs/2411.08236} {2411.08236} \BibitemShut
  {NoStop}%
\bibitem [{\citenamefont {Sevinçli}\ and\ \citenamefont
  {Pohl}(2014)}]{Sevinçli.2014}%
  \BibitemOpen
  \bibfield  {author} {\bibinfo {author} {\bibfnamefont {S.}~\bibnamefont
  {Sevinçli}}\ and\ \bibinfo {author} {\bibfnamefont {T.}~\bibnamefont
  {Pohl}},\ }\bibfield  {title} {\bibinfo {title} {{Microwave control of
  Rydberg atom interactions}},\ }\href
  {https://doi.org/10.1088/1367-2630/16/12/123036} {\bibfield  {journal}
  {\bibinfo  {journal} {New Journal of Physics}\ }\textbf {\bibinfo {volume}
  {16}},\ \bibinfo {pages} {123036} (\bibinfo {year} {2014})},\ \Eprint
  {https://arxiv.org/abs/1412.4925} {1412.4925} \BibitemShut {NoStop}%
\bibitem [{\citenamefont {Shi}\ and\ \citenamefont {Kennedy}(2017)}]{Shi.2017}%
  \BibitemOpen
  \bibfield  {author} {\bibinfo {author} {\bibfnamefont {X.-F.}\ \bibnamefont
  {Shi}}\ and\ \bibinfo {author} {\bibfnamefont {T.~A.~B.}\ \bibnamefont
  {Kennedy}},\ }\bibfield  {title} {\bibinfo {title} {{Annulled van der Waals
  interaction and fast Rydberg quantum gates}},\ }\href
  {https://doi.org/10.1103/physreva.95.043429} {\bibfield  {journal} {\bibinfo
  {journal} {Physical Review A}\ }\textbf {\bibinfo {volume} {95}},\ \bibinfo
  {pages} {043429} (\bibinfo {year} {2017})},\ \Eprint
  {https://arxiv.org/abs/1606.08516} {1606.08516} \BibitemShut {NoStop}%
\bibitem [{\citenamefont {Young}\ \emph {et~al.}(2021)\citenamefont {Young},
  \citenamefont {Bienias}, \citenamefont {Belyansky}, \citenamefont {Kaufman},\
  and\ \citenamefont {Gorshkov}}]{Young.2021}%
  \BibitemOpen
  \bibfield  {author} {\bibinfo {author} {\bibfnamefont {J.~T.}\ \bibnamefont
  {Young}}, \bibinfo {author} {\bibfnamefont {P.}~\bibnamefont {Bienias}},
  \bibinfo {author} {\bibfnamefont {R.}~\bibnamefont {Belyansky}}, \bibinfo
  {author} {\bibfnamefont {A.~M.}\ \bibnamefont {Kaufman}},\ and\ \bibinfo
  {author} {\bibfnamefont {A.~V.}\ \bibnamefont {Gorshkov}},\ }\bibfield
  {title} {\bibinfo {title} {{Asymmetric Blockade and Multiqubit Gates via
  Dipole-Dipole Interactions}},\ }\href
  {https://doi.org/10.1103/physrevlett.127.120501} {\bibfield  {journal}
  {\bibinfo  {journal} {Physical Review Letters}\ }\textbf {\bibinfo {volume}
  {127}},\ \bibinfo {pages} {120501} (\bibinfo {year} {2021})},\ \Eprint
  {https://arxiv.org/abs/2006.02486} {2006.02486} \BibitemShut {NoStop}%
\bibitem [{\citenamefont {Yuan}\ \emph {et~al.}(2023)\citenamefont {Yuan},
  \citenamefont {Zhang}, \citenamefont {Bigagli}, \citenamefont {Warner},
  \citenamefont {Stevenson},\ and\ \citenamefont {Will}}]{Yuan.2023}%
  \BibitemOpen
  \bibfield  {author} {\bibinfo {author} {\bibfnamefont {W.}~\bibnamefont
  {Yuan}}, \bibinfo {author} {\bibfnamefont {S.}~\bibnamefont {Zhang}},
  \bibinfo {author} {\bibfnamefont {N.}~\bibnamefont {Bigagli}}, \bibinfo
  {author} {\bibfnamefont {C.}~\bibnamefont {Warner}}, \bibinfo {author}
  {\bibfnamefont {I.}~\bibnamefont {Stevenson}},\ and\ \bibinfo {author}
  {\bibfnamefont {S.}~\bibnamefont {Will}},\ }\bibfield  {title} {\bibinfo
  {title} {{A planar cloverleaf antenna for circularly polarized microwave
  fields in atomic and molecular physics experiments}},\ }\href
  {https://doi.org/10.1063/5.0167572} {\bibfield  {journal} {\bibinfo
  {journal} {Review of Scientific Instruments}\ }\textbf {\bibinfo {volume}
  {94}},\ \bibinfo {pages} {123201} (\bibinfo {year} {2023})},\ \Eprint
  {https://arxiv.org/abs/2306.14791} {2306.14791} \BibitemShut {NoStop}%
\bibitem [{\citenamefont {Bigagli}\ \emph {et~al.}(2023)\citenamefont
  {Bigagli}, \citenamefont {Warner}, \citenamefont {Yuan}, \citenamefont
  {Zhang}, \citenamefont {Stevenson}, \citenamefont {Karman},\ and\
  \citenamefont {Will}}]{Bigagli.2023}%
  \BibitemOpen
  \bibfield  {author} {\bibinfo {author} {\bibfnamefont {N.}~\bibnamefont
  {Bigagli}}, \bibinfo {author} {\bibfnamefont {C.}~\bibnamefont {Warner}},
  \bibinfo {author} {\bibfnamefont {W.}~\bibnamefont {Yuan}}, \bibinfo {author}
  {\bibfnamefont {S.}~\bibnamefont {Zhang}}, \bibinfo {author} {\bibfnamefont
  {I.}~\bibnamefont {Stevenson}}, \bibinfo {author} {\bibfnamefont
  {T.}~\bibnamefont {Karman}},\ and\ \bibinfo {author} {\bibfnamefont
  {S.}~\bibnamefont {Will}},\ }\bibfield  {title} {\bibinfo {title}
  {{Collisionally stable gas of bosonic dipolar ground-state molecules}},\
  }\href {https://doi.org/10.1038/s41567-023-02200-6} {\bibfield  {journal}
  {\bibinfo  {journal} {Nature Physics}\ }\textbf {\bibinfo {volume} {19}},\
  \bibinfo {pages} {1579} (\bibinfo {year} {2023})},\ \Eprint
  {https://arxiv.org/abs/2303.16845} {2303.16845} \BibitemShut {NoStop}%
\bibitem [{\citenamefont {Lin}\ \emph {et~al.}(2023)\citenamefont {Lin},
  \citenamefont {Chen}, \citenamefont {Jin}, \citenamefont {Shi}, \citenamefont
  {Deng}, \citenamefont {Zhang}, \citenamefont {Quéméner}, \citenamefont
  {Shi}, \citenamefont {Yi},\ and\ \citenamefont {Wang}}]{Lin.2023}%
  \BibitemOpen
  \bibfield  {author} {\bibinfo {author} {\bibfnamefont {J.}~\bibnamefont
  {Lin}}, \bibinfo {author} {\bibfnamefont {G.}~\bibnamefont {Chen}}, \bibinfo
  {author} {\bibfnamefont {M.}~\bibnamefont {Jin}}, \bibinfo {author}
  {\bibfnamefont {Z.}~\bibnamefont {Shi}}, \bibinfo {author} {\bibfnamefont
  {F.}~\bibnamefont {Deng}}, \bibinfo {author} {\bibfnamefont {W.}~\bibnamefont
  {Zhang}}, \bibinfo {author} {\bibfnamefont {G.}~\bibnamefont {Quéméner}},
  \bibinfo {author} {\bibfnamefont {T.}~\bibnamefont {Shi}}, \bibinfo {author}
  {\bibfnamefont {S.}~\bibnamefont {Yi}},\ and\ \bibinfo {author}
  {\bibfnamefont {D.}~\bibnamefont {Wang}},\ }\bibfield  {title} {\bibinfo
  {title} {{Microwave Shielding of Bosonic NaRb Molecules}},\ }\href
  {https://doi.org/10.1103/physrevx.13.031032} {\bibfield  {journal} {\bibinfo
  {journal} {Physical Review X}\ }\textbf {\bibinfo {volume} {13}},\ \bibinfo
  {pages} {031032} (\bibinfo {year} {2023})},\ \Eprint
  {https://arxiv.org/abs/2304.08312} {2304.08312} \BibitemShut {NoStop}%
\bibitem [{\citenamefont {Sedlacek}\ \emph {et~al.}(2013)\citenamefont
  {Sedlacek}, \citenamefont {Schwettmann}, \citenamefont {Kübler},\ and\
  \citenamefont {Shaffer}}]{Sedlacek.2013}%
  \BibitemOpen
  \bibfield  {author} {\bibinfo {author} {\bibfnamefont {J.~A.}\ \bibnamefont
  {Sedlacek}}, \bibinfo {author} {\bibfnamefont {A.}~\bibnamefont
  {Schwettmann}}, \bibinfo {author} {\bibfnamefont {H.}~\bibnamefont
  {Kübler}},\ and\ \bibinfo {author} {\bibfnamefont {J.~P.}\ \bibnamefont
  {Shaffer}},\ }\bibfield  {title} {\bibinfo {title} {{Atom-Based Vector
  Microwave Electrometry Using Rubidium Rydberg Atoms in a Vapor Cell}},\
  }\href {https://doi.org/10.1103/physrevlett.111.063001} {\bibfield  {journal}
  {\bibinfo  {journal} {Physical Review Letters}\ }\textbf {\bibinfo {volume}
  {111}},\ \bibinfo {pages} {063001} (\bibinfo {year} {2013})},\ \Eprint
  {https://arxiv.org/abs/1304.4299} {1304.4299} \BibitemShut {NoStop}%
\bibitem [{\citenamefont {Cloutman}\ \emph {et~al.}(2024)\citenamefont
  {Cloutman}, \citenamefont {Chilcott}, \citenamefont {Elliott}, \citenamefont
  {Otto}, \citenamefont {Deb},\ and\ \citenamefont
  {Kjærgaard}}]{Cloutman.2024}%
  \BibitemOpen
  \bibfield  {author} {\bibinfo {author} {\bibfnamefont {M.}~\bibnamefont
  {Cloutman}}, \bibinfo {author} {\bibfnamefont {M.}~\bibnamefont {Chilcott}},
  \bibinfo {author} {\bibfnamefont {A.}~\bibnamefont {Elliott}}, \bibinfo
  {author} {\bibfnamefont {J.~S.}\ \bibnamefont {Otto}}, \bibinfo {author}
  {\bibfnamefont {A.~B.}\ \bibnamefont {Deb}},\ and\ \bibinfo {author}
  {\bibfnamefont {N.}~\bibnamefont {Kjærgaard}},\ }\bibfield  {title}
  {\bibinfo {title} {{Polarization-insensitive microwave electrometry using
  Rydberg atoms}},\ }\href {https://doi.org/10.1103/physrevapplied.21.044025}
  {\bibfield  {journal} {\bibinfo  {journal} {Physical Review Applied}\
  }\textbf {\bibinfo {volume} {21}},\ \bibinfo {pages} {044025} (\bibinfo
  {year} {2024})},\ \Eprint {https://arxiv.org/abs/2312.01974} {2312.01974}
  \BibitemShut {NoStop}%
\bibitem [{\citenamefont {You}\ \emph {et~al.}(2024)\citenamefont {You},
  \citenamefont {Cai}, \citenamefont {Zhang}, \citenamefont {Song},\ and\
  \citenamefont {Liu}}]{You.2024}%
  \BibitemOpen
  \bibfield  {author} {\bibinfo {author} {\bibfnamefont {S.~H.}\ \bibnamefont
  {You}}, \bibinfo {author} {\bibfnamefont {M.~H.}\ \bibnamefont {Cai}},
  \bibinfo {author} {\bibfnamefont {H.~A.}\ \bibnamefont {Zhang}}, \bibinfo
  {author} {\bibfnamefont {Z.~F.}\ \bibnamefont {Song}},\ and\ \bibinfo
  {author} {\bibfnamefont {H.}~\bibnamefont {Liu}},\ }\bibfield  {title}
  {\bibinfo {title} {{RF spectra induced by different polarized microwave}},\
  }\href {https://doi.org/10.1063/5.0173545} {\bibfield  {journal} {\bibinfo
  {journal} {AIP Advances}\ }\textbf {\bibinfo {volume} {14}},\ \bibinfo
  {pages} {015245} (\bibinfo {year} {2024})}\BibitemShut {NoStop}%
\bibitem [{\citenamefont {Cloutman}\ \emph {et~al.}(2025)\citenamefont
  {Cloutman}, \citenamefont {Chilcott}, \citenamefont {Elliott}, \citenamefont
  {Otto}, \citenamefont {Deb},\ and\ \citenamefont
  {Kjærgaard}}]{Cloutman.2025}%
  \BibitemOpen
  \bibfield  {author} {\bibinfo {author} {\bibfnamefont {M.}~\bibnamefont
  {Cloutman}}, \bibinfo {author} {\bibfnamefont {M.}~\bibnamefont {Chilcott}},
  \bibinfo {author} {\bibfnamefont {A.}~\bibnamefont {Elliott}}, \bibinfo
  {author} {\bibfnamefont {J.~S.}\ \bibnamefont {Otto}}, \bibinfo {author}
  {\bibfnamefont {A.~B.}\ \bibnamefont {Deb}},\ and\ \bibinfo {author}
  {\bibfnamefont {N.}~\bibnamefont {Kjærgaard}},\ }\bibfield  {title}
  {\bibinfo {title} {{Quantum-enabled Rydberg atomic polarimetry of
  radio-frequency fields}},\ }\bibfield  {journal} {\bibinfo  {journal}
  {arXiv}\ }\href {https://doi.org/10.48550/arxiv.2503.17997}
  {10.48550/arxiv.2503.17997} (\bibinfo {year} {2025}),\ \Eprint
  {https://arxiv.org/abs/2503.17997} {2503.17997} \BibitemShut {NoStop}%
\bibitem [{\citenamefont {Sheng-yuan}\ \emph {et~al.}(2022)\citenamefont
  {Sheng-yuan}, \citenamefont {Ming-yong}, \citenamefont {Hao}, \citenamefont
  {Xiao-bo},\ and\ \citenamefont {Lin-jie}}]{Ren.2022}%
  \BibitemOpen
  \bibfield  {author} {\bibinfo {author} {\bibfnamefont {R.}~\bibnamefont
  {Sheng-yuan}}, \bibinfo {author} {\bibfnamefont {J.}~\bibnamefont
  {Ming-yong}}, \bibinfo {author} {\bibfnamefont {Z.}~\bibnamefont {Hao}},
  \bibinfo {author} {\bibfnamefont {W.}~\bibnamefont {Xiao-bo}},\ and\ \bibinfo
  {author} {\bibfnamefont {Z.}~\bibnamefont {Lin-jie}},\ }\bibfield  {title}
  {\bibinfo {title} {{Atom-Based Vector Measurement of Near Field Scattering
  Field of Radio Frequency Identification Tag}},\ }\href
  {https://doi.org/10.3964/j.issn.1000-0593(2022)01-0298-06} {\bibfield
  {journal} {\bibinfo  {journal} {Spectroscopy and Spectral Analysis}\ }\textbf
  {\bibinfo {volume} {42}},\ \bibinfo {pages} {298} (\bibinfo {year}
  {2022})}\BibitemShut {NoStop}%
\bibitem [{\citenamefont {Bai}\ \emph {et~al.}(2019)\citenamefont {Bai},
  \citenamefont {Fan}, \citenamefont {Hao}, \citenamefont {Spong},
  \citenamefont {Jiao},\ and\ \citenamefont {Zhao}}]{Bai.2019}%
  \BibitemOpen
  \bibfield  {author} {\bibinfo {author} {\bibfnamefont {J.}~\bibnamefont
  {Bai}}, \bibinfo {author} {\bibfnamefont {J.}~\bibnamefont {Fan}}, \bibinfo
  {author} {\bibfnamefont {L.}~\bibnamefont {Hao}}, \bibinfo {author}
  {\bibfnamefont {N.~L.~R.}\ \bibnamefont {Spong}}, \bibinfo {author}
  {\bibfnamefont {Y.}~\bibnamefont {Jiao}},\ and\ \bibinfo {author}
  {\bibfnamefont {J.}~\bibnamefont {Zhao}},\ }\bibfield  {title} {\bibinfo
  {title} {{Measurement of the Near Field Distribution of a Microwave Horn
  Using a Resonant Atomic Probe}},\ }\href {https://doi.org/10.3390/app9224895}
  {\bibfield  {journal} {\bibinfo  {journal} {Applied Sciences}\ }\textbf
  {\bibinfo {volume} {9}},\ \bibinfo {pages} {4895} (\bibinfo {year}
  {2019})}\BibitemShut {NoStop}%
\bibitem [{\citenamefont {Elgee}\ \emph {et~al.}(2024)\citenamefont {Elgee},
  \citenamefont {Cox}, \citenamefont {Hill}, \citenamefont {Kunz},\ and\
  \citenamefont {Meyer}}]{Elgee.2024}%
  \BibitemOpen
  \bibfield  {author} {\bibinfo {author} {\bibfnamefont {P.~K.}\ \bibnamefont
  {Elgee}}, \bibinfo {author} {\bibfnamefont {K.~C.}\ \bibnamefont {Cox}},
  \bibinfo {author} {\bibfnamefont {J.~C.}\ \bibnamefont {Hill}}, \bibinfo
  {author} {\bibfnamefont {P.~D.}\ \bibnamefont {Kunz}},\ and\ \bibinfo
  {author} {\bibfnamefont {D.~H.}\ \bibnamefont {Meyer}},\ }\bibfield  {title}
  {\bibinfo {title} {{Complete three-dimensional vector polarimetry with a
  Rydberg-atom rf electrometer}},\ }\href
  {https://doi.org/10.1103/physrevapplied.22.064012} {\bibfield  {journal}
  {\bibinfo  {journal} {Physical Review Applied}\ }\textbf {\bibinfo {volume}
  {22}},\ \bibinfo {pages} {064012} (\bibinfo {year} {2024})},\ \Eprint
  {https://arxiv.org/abs/2407.20369} {2407.20369} \BibitemShut {NoStop}%
\bibitem [{\citenamefont {Wang}\ \emph {et~al.}(2023)\citenamefont {Wang},
  \citenamefont {Jia}, \citenamefont {Hao}, \citenamefont {Cui}, \citenamefont
  {Zhou}, \citenamefont {Liu}, \citenamefont {Mei}, \citenamefont {Yu},
  \citenamefont {Liu}, \citenamefont {Zhang}, \citenamefont {Xie},\ and\
  \citenamefont {Zhong}}]{Wang.2023}%
  \BibitemOpen
  \bibfield  {author} {\bibinfo {author} {\bibfnamefont {Y.}~\bibnamefont
  {Wang}}, \bibinfo {author} {\bibfnamefont {F.}~\bibnamefont {Jia}}, \bibinfo
  {author} {\bibfnamefont {J.}~\bibnamefont {Hao}}, \bibinfo {author}
  {\bibfnamefont {Y.}~\bibnamefont {Cui}}, \bibinfo {author} {\bibfnamefont
  {F.}~\bibnamefont {Zhou}}, \bibinfo {author} {\bibfnamefont {X.}~\bibnamefont
  {Liu}}, \bibinfo {author} {\bibfnamefont {J.}~\bibnamefont {Mei}}, \bibinfo
  {author} {\bibfnamefont {Y.}~\bibnamefont {Yu}}, \bibinfo {author}
  {\bibfnamefont {Y.}~\bibnamefont {Liu}}, \bibinfo {author} {\bibfnamefont
  {J.}~\bibnamefont {Zhang}}, \bibinfo {author} {\bibfnamefont
  {F.}~\bibnamefont {Xie}},\ and\ \bibinfo {author} {\bibfnamefont
  {Z.}~\bibnamefont {Zhong}},\ }\bibfield  {title} {\bibinfo {title} {{Precise
  measurement of microwave polarization using a Rydberg atom-based mixer}},\
  }\href {https://doi.org/10.1364/oe.485662} {\bibfield  {journal} {\bibinfo
  {journal} {Optics Express}\ }\textbf {\bibinfo {volume} {31}},\ \bibinfo
  {pages} {10449} (\bibinfo {year} {2023})}\BibitemShut {NoStop}%
\bibitem [{\citenamefont {Yin}\ \emph {et~al.}(2024)\citenamefont {Yin},
  \citenamefont {Zhang}, \citenamefont {Jia}, \citenamefont {Wang},
  \citenamefont {Wang}, \citenamefont {Hao}, \citenamefont {Cui}, \citenamefont
  {Liu},\ and\ \citenamefont {Zhong}}]{Yin.2024}%
  \BibitemOpen
  \bibfield  {author} {\bibinfo {author} {\bibfnamefont {W.}~\bibnamefont
  {Yin}}, \bibinfo {author} {\bibfnamefont {J.}~\bibnamefont {Zhang}}, \bibinfo
  {author} {\bibfnamefont {F.}~\bibnamefont {Jia}}, \bibinfo {author}
  {\bibfnamefont {Y.}~\bibnamefont {Wang}}, \bibinfo {author} {\bibfnamefont
  {Y.}~\bibnamefont {Wang}}, \bibinfo {author} {\bibfnamefont {J.}~\bibnamefont
  {Hao}}, \bibinfo {author} {\bibfnamefont {Y.}~\bibnamefont {Cui}}, \bibinfo
  {author} {\bibfnamefont {Y.}~\bibnamefont {Liu}},\ and\ \bibinfo {author}
  {\bibfnamefont {Z.}~\bibnamefont {Zhong}},\ }\bibfield  {title} {\bibinfo
  {title} {{Measurement of microwave polarization using two polarization
  orthogonal local microwave electric fields in a Rydberg atom-based mixer}},\
  }\href {https://doi.org/10.1364/oe.538440} {\bibfield  {journal} {\bibinfo
  {journal} {Optics Express}\ }\textbf {\bibinfo {volume} {32}},\ \bibinfo
  {pages} {38372} (\bibinfo {year} {2024})}\BibitemShut {NoStop}%
\bibitem [{\citenamefont {Fleischhauer}\ \emph {et~al.}(2005)\citenamefont
  {Fleischhauer}, \citenamefont {Imamoglu},\ and\ \citenamefont
  {Marangos}}]{Fleischhauer.2005}%
  \BibitemOpen
  \bibfield  {author} {\bibinfo {author} {\bibfnamefont {M.}~\bibnamefont
  {Fleischhauer}}, \bibinfo {author} {\bibfnamefont {A.}~\bibnamefont
  {Imamoglu}},\ and\ \bibinfo {author} {\bibfnamefont {J.~P.}\ \bibnamefont
  {Marangos}},\ }\bibfield  {title} {\bibinfo {title} {{Electromagnetically
  induced transparency: Optics in coherent media}},\ }\href
  {https://doi.org/10.1103/revmodphys.77.633} {\bibfield  {journal} {\bibinfo
  {journal} {Reviews of Modern Physics}\ }\textbf {\bibinfo {volume} {77}},\
  \bibinfo {pages} {633} (\bibinfo {year} {2005})}\BibitemShut {NoStop}%
\bibitem [{\citenamefont {Shi}\ \emph {et~al.}(2016)\citenamefont {Shi},
  \citenamefont {Svetlichnyy},\ and\ \citenamefont {Kennedy}}]{Shi.2016}%
  \BibitemOpen
  \bibfield  {author} {\bibinfo {author} {\bibfnamefont {X.-F.}\ \bibnamefont
  {Shi}}, \bibinfo {author} {\bibfnamefont {P.}~\bibnamefont {Svetlichnyy}},\
  and\ \bibinfo {author} {\bibfnamefont {T.~A.~B.}\ \bibnamefont {Kennedy}},\
  }\bibfield  {title} {\bibinfo {title} {{Spin–charge separation of
  dark-state polaritons in a Rydberg medium}},\ }\href
  {https://doi.org/10.1088/0953-4075/49/7/074005} {\bibfield  {journal}
  {\bibinfo  {journal} {Journal of Physics B: Atomic, Molecular and Optical
  Physics}\ }\textbf {\bibinfo {volume} {49}},\ \bibinfo {pages} {074005}
  (\bibinfo {year} {2016})}\BibitemShut {NoStop}%
\bibitem [{\citenamefont {Liu}\ \emph {et~al.}(2022)\citenamefont {Liu},
  \citenamefont {Liao}, \citenamefont {Zhang}, \citenamefont {Tu},
  \citenamefont {Bian}, \citenamefont {Li}, \citenamefont {Zheng},
  \citenamefont {Li}, \citenamefont {Huang}, \citenamefont {Yan},\ and\
  \citenamefont {Zhu}}]{Liu.20228js}%
  \BibitemOpen
  \bibfield  {author} {\bibinfo {author} {\bibfnamefont {X.-H.}\ \bibnamefont
  {Liu}}, \bibinfo {author} {\bibfnamefont {K.-Y.}\ \bibnamefont {Liao}},
  \bibinfo {author} {\bibfnamefont {Z.-X.}\ \bibnamefont {Zhang}}, \bibinfo
  {author} {\bibfnamefont {H.-T.}\ \bibnamefont {Tu}}, \bibinfo {author}
  {\bibfnamefont {W.}~\bibnamefont {Bian}}, \bibinfo {author} {\bibfnamefont
  {Z.-Q.}\ \bibnamefont {Li}}, \bibinfo {author} {\bibfnamefont {S.-Y.}\
  \bibnamefont {Zheng}}, \bibinfo {author} {\bibfnamefont {H.-H.}\ \bibnamefont
  {Li}}, \bibinfo {author} {\bibfnamefont {W.}~\bibnamefont {Huang}}, \bibinfo
  {author} {\bibfnamefont {H.}~\bibnamefont {Yan}},\ and\ \bibinfo {author}
  {\bibfnamefont {S.-L.}\ \bibnamefont {Zhu}},\ }\bibfield  {title} {\bibinfo
  {title} {{Continuous-Frequency Microwave Heterodyne Detection in an Atomic
  Vapor Cell}},\ }\href {https://doi.org/10.1103/physrevapplied.18.054003}
  {\bibfield  {journal} {\bibinfo  {journal} {Physical Review Applied}\
  }\textbf {\bibinfo {volume} {18}},\ \bibinfo {pages} {054003} (\bibinfo
  {year} {2022})}\BibitemShut {NoStop}%
\bibitem [{\citenamefont {Kurdak}\ \emph
  {et~al.}(2025{\natexlab{b}})\citenamefont {Kurdak}, \citenamefont {Li},
  \citenamefont {Banner}, \citenamefont {Porto},\ and\ \citenamefont
  {Rolston}}]{Zenodo.2025}%
  \BibitemOpen
  \bibfield  {author} {\bibinfo {author} {\bibfnamefont {D.}~\bibnamefont
  {Kurdak}}, \bibinfo {author} {\bibfnamefont {Y.}~\bibnamefont {Li}}, \bibinfo
  {author} {\bibfnamefont {P.~R.}\ \bibnamefont {Banner}}, \bibinfo {author}
  {\bibfnamefont {J.~V.}\ \bibnamefont {Porto}},\ and\ \bibinfo {author}
  {\bibfnamefont {S.~L.}\ \bibnamefont {Rolston}},\ }\href
  {https://doi.org/10.5281/zenodo.16783939} {\bibinfo {title} {{Datasets for
  ``High-Fidelity Microwave-Polarization Control in a Rydberg-Ensemble
  Experiment''}}},\ \bibinfo {howpublished} {Dataset on Zenodo} (\bibinfo
  {year} {2025}{\natexlab{b}})\BibitemShut {NoStop}%
\end{thebibliography}
\end{document}